\documentclass[a4paper]{jpconf}
\usepackage{graphicx}

\graphicspath{ {/} }

\usepackage{hyperref}
\hypersetup{
	hyperindex,
	breaklinks,
	colorlinks=true,
	linkcolor=blue,
	citecolor=purple,
	bookmarks=true,
	bookmarksopen=true,
	bookmarksopenlevel=2,
	pdfstartpage={1},
	pdfstartview={FitH},
	pdfview={FitH 0},
	pdfauthor={N C Constantinou, A Lozano-Duran, M-A Nikolaidis, B F Farrell, P J Ioannou and J Jimenez},
	pdftitle={Turbulence in the highly restricted dynamics of a closure at second order: comparison with DNS},
 }

\usepackage[sort&compress]{natbib}
\bibpunct{[}{]}{,}{n}{}{,}

\usepackage{microtype}

\usepackage{bm}
\usepackage{subfigure}

\usepackage{amsmath,amssymb}
\usepackage[usenames,dvipsnames]{xcolor}

\newcommand{\dispdot}[2][.2ex]{\dot{\raisebox{0pt}[\dimexpr\height+#1][\depth]{$#2$}}}

\def\U{\bm{\mathsf{U}}}

\newcounter{saveeqn}%

\newcommand{\be}{\begin{equation}}
\newcommand{\ee}{\end{equation}}
\newcommand{\bdm}{\begin{equation*}}
\newcommand{\edm}{\end{equation*}}
\newcommand{\bea}{\begin{eqnarray}}
\newcommand{\eea}{\end{eqnarray}}

\newcommand{\partialf}[2]
{
 \ifthenelse{\equal{#1}{}}{\frac{\partial}{\partial #2}}{\frac{\partial #1}{\partial #2}}
}

\newcommand{\sgn}{\mathop{\mathrm{sgn}}}

\renewcommand{\(}{\left(}
\renewcommand{\)}{\right)}
\renewcommand{\[}{\left[}
\renewcommand{\]}{\right]}

\newcommand{\Del}{\Delta}

\newcommand{\df}{\textrm{d}}

\newcommand{\la}{\lambda}

\newsavebox{\astrutbox}
\sbox{\astrutbox}{\rule[-5pt]{0pt}{20pt}}

\usepackage{upgreek}

\def\bit{\vphantom{\dot{W}}}

\renewcommand{\Re}{\textrm{Re}}
\newcommand{\Ret}{\textrm{Re}_\tau}
\newcommand{\ut}{u_\tau}

\renewcommand{\U}{\mathbf{U}}
\renewcommand{\u}{\mathbf{u}}

\begin{document}
\title{Turbulence in the highly restricted dynamics of a closure at second order: comparison with DNS}

\author{N~C~Constantinou$^1$, A~Lozano-Dur\'an$^2$,  M-A~Nikolaidis$^1$, B~F~Farrell$^3$, P~J~Ioannou$^1$,  J~Jim\'enez$^2$}

\address{$^1$University of Athens, Department of Physics, Panepistimiopolis, Zografos, 15784 Athens, Greece}
\address{$^2$School of Aeronautics, Universidad Polit\'ecnica de Madrid, 28040 Madrid, Spain}
\address{$^3$Department of Earth and Planetary Sciences, Harvard University, Cambridge, MA 02138, USA}

\ead{\href{mailto:navidcon@phys.uoa.gr}{navidcon@phys.uoa.gr}}

\begin{abstract}
S3T (Stochastic Structural Stability Theory) employs a closure at second order  to obtain the  dynamics of the statistical mean turbulent state. When S3T is implemented as a coupled set of equations for the streamwise mean and perturbation states, nonlinearity in the dynamics is restricted to interaction between the mean and perturbations. 
The S3T statistical mean state dynamics can be approximately implemented by similarly restricting the dynamics used in a direct numerical simulation (DNS) of the full Navier-Stokes equations (referred to as the NS system). Although this restricted nonlinear system (referred to as the RNL system) is greatly simplified in its dynamics in comparison to the associated NS,  it nevertheless self-sustains a turbulent state in wall-bounded shear flow with structures and dynamics comparable to that  in observed turbulence. 
Moreover, RNL turbulence can be analyzed effectively using  theoretical methods developed to study the closely related S3T system. 
In order to better understand  RNL turbulence and its relation to  NS turbulence, an extensive comparison is made of diagnostics of structure and dynamics in these systems.
Although quantitative differences are found, the results show that turbulence in the  RNL system closely parallels that in NS and suggest that the S3T/RNL system provides a promising reduced complexity model for studying turbulence in wall-bounded shear flows.
\end{abstract}

\section{Introduction}

The Navier-Stokes equations (NS), while comprising the complete dynamics of turbulence, have at least two disadvantages for theoretical investigation of the physics of turbulence: NS lacks analytical solution for the case of the fully turbulent state and the nonlinear advection term results in turbulent  states of high complexity which tends to obscure the  fundamental mechanisms underlying the turbulence. One approach to overcoming these impediments has been the search for simplifications of NS that retain essential features of the turbulence dynamics. The Linearized Navier-Stokes equations (LNS) provide one example of the successful application of this approach in which the power of linear systems theory is made available to the study of turbulence~\cite{Farrell-Ioannou-1996a,Farrell-Ioannou-1996b}. The LNS system captures the non-normal mechanism responsible for perturbation growth in NS~\cite{Henningson-1996b}. This linear mechanism retained in LNS underlies both the process of subcritical transition to turbulence and the maintenance of the turbulent state~\cite{Butler-Farrell-1992,Henningson-Reddy-1994,
Farrell-Ioannou-1994b,Farrell-Ioannou-2012}. However, linear models are unable to capture other essential phenomena in turbulence that are intrinsically nonlinear including establishment of the turbulent mean velocity profile and maintenance of turbulence in a statistically steady state in the absence of extrinsic excitation, the mechanism of which is referred to as the self-sustaining process (SSP). 
While there remain many further aspects of the turbulent state to be addressed, such as the structures and spectrum of the inertial subrange, 
a basis of understanding of wall turbulence comprises these fundamental aspects. One approach to studying these phenomena that has proven successful 
is to impose simplifications that isolate certain mechanisms of the turbulence dynamics. 
An example is  restricting the  size of the channel~\citep{Jimenez-Moin-1991}. 
The existence  of the minimal channel for sustaining turbulence isolated  the streak structure component of the SSP\footnote{In this context the term ``streak''  describes well-defined elongated regions of spanwise alternating bands of low and high speed fluid superimposed on the mean shear.} and showed that the SSP cycle comprises  interaction among the streak, 
a  pair of streamwise vortices of opposite sign (referred to as a rolls) and the perturbation field. 
In minimal channel turbulence  a 
single streak fits in the channel 
and the streamwise extent of the channel is of such length  that  the perturbations can
adopt a  structure allowing them to extract energy from the mean flow 
through a time dependent interaction~\citep{Jimenez-Moin-1991,Hamilton-etal-1995,Jimenez-Pinelli-1999}.
Another example is imposing simplification of the dynamics by isolating specific regions of the turbulent flow. 
For example, it was found that when the inner and outer regions of the flow are isolated the turbulence was independently 
maintained in each region  arguing that an independent SSP is operating in these regions~\citep{Jimenez-Pinelli-1999,Tuerke-Jimenez-2013,Mizuno-Jimenez-2013}.

The minimal channel flow studies referred to above focused attention on the effort to obtain a self-consistent 
dynamical description of the interaction among the streak, the roll and the perturbation field.
The streaks in minimal simulations  are associated with the zero streamwise harmonic ($k_x=0$). 
This observation motivates constructing a reduced complexity turbulence model based on the simplest closure of the  LNS equations, which govern 
interaction between the streamwise mean flow (with the  streak  structure included) and the streamwise varying perturbation field (characterized by the $k_x\ne0$ flow field).
This system is obtained by augmenting LNS by only the nonlinearity resulting from including
the feedback of the $k_x\ne0$  perturbations on the $k_x=0$  streamwise mean flow.
This dynamics, which greatly restricts the nonlinearity of the NS, 
will be referred to as the restricted nonlinear system (RNL).
Remarkably, as will be demonstrated in this work,
the RNL self-sustains a realistic turbulent state, not only in minimal channels at low Reynolds numbers, but also in large channels and at 
moderate Reynolds numbers.

The RNL system was introduced here as a simple extension of LNS that supports self-sustained turbulence.
However, the RNL has deeper roots: it approximates the
second order cumulant closure of the NS, which is the basis of the Stochastic Structural Stability Theory (S3T).  S3T defines a statistical mean state dynamical system and implementations of this system have  recently been used to  develop
theories of turbulence~\citep{Farrell-Ioannou-2003-structural,Farrell-Ioannou-2007-structure,
Marston-etal-2008,Tobias-etal-2011,Srinivasan-Young-2012,Bakas-Ioannou-2012-prl, Constantinou-etal-2013}.
S3T employs an ensemble closure which produces autonomous statistical mean state 
dynamics in which turbulent  mean states exist as statistical equilibria. This makes the turbulent state available as an 
object for stability study,  extending  classical hydrodynamic stability  theory which  addresses 
only the stability of stationary sample state solutions of the NS. S3T, in contrast, can be used to determine the structural stability of  statistical mean state entities such as attractors (characterised by a specific probability density function), so that for instance, when such an attractor  becomes S3T unstable the fluid state bifurcates to a new attractor characterized by a different probability density function. An example application of S3T is to a constant shear flow subjected to homogeneous turbulence excitation in which it is found that a bifurcation occurs when the Reynolds number exceeds a critical value resulting in  a new stationary state  comprising the mean flow with a roll-streak structure and a perturbation field supporting it in a new statistical steady state~\cite{Farrell-Ioannou-2012}. At even higher Reynolds number a saddle-node bifurcation occurs  in the S3T system and the  flow transitions  to a time dependent state that self-sustains, which is identified with transition to the turbulent state.
In the self-sustaining turbulent state the statistical mean state dynamics of  S3T approaches  the deterministic dynamics of the RNL.
Recent work has exploited S3T dynamics to explore the role of streamwise coherent structures  in turbulence, including the dynamics of the roll and streak structures~\cite{Farrell-Ioannou-2012-CTRv2,Thomas-etal-2013-sustain}. In this paper, we verify predictions of  S3T for  turbulence structure  at high Reynolds numbers in pressure driven channel flow, by comparing flow statistics, structures, and dynamical diagnostics   obtained from the RNL system to results obtained in the S3T system and  to direct numerical simulations  (DNS) of the full NS equations.

\section{Modeling framework}
\label{sec:framework}

Consider a pressure driven plane Poiseuille flow maintained by application of the time-dependent pressure, $G(t)x$, where $x$ is the streamwise coordinate.  
The  wall-normal direction is $y$ and the spanwise direction is $z$. 
The  lengths of the channel in the streamwise, wall-normal  and 
spanwise  direction are respectively $  L_x$, $2h$ and $L_z$. The channel walls are at $y/h=0$~and~$2$.  
Streamwise mean, spanwise mean and time mean   quantities
are denoted respectively by an overbar, $\overline{\,\vphantom{W}\bullet\,} 
= L_x^{-1} \int_0^{ L_x} \bullet ~ \df x $, square brackets, $[\,\bullet\,]=L_z^{-1} \int_0^{L_z} \bullet ~\df z$ and  a  wide hat $\widehat{\;\bullet\;} = T^{-1} \int_0^T \bullet~\df t$, with  $T$ sufficiently long.   
The velocity, $\mathbf{u}$, is decomposed  into its streamwise mean value, denoted 
$\U(y,z,t)$, and the deviation from the mean (the perturbation), $\u'(x,y,z,t)$, so that the flow velocity is
$\mathbf{u}=\mathbf{U}+\mathbf{u}'$. The pressure gradient is similarly written as
$\nabla p= \nabla\(-G(t)  x + P(y,z,t)+ p'(x,y,z,t) \)$.  
The NS can be then decomposed into an equation for the  mean and an equation for the perturbation as follows:   
\begin{subequations}
\begin{align}
\partial_t\mathbf{U}&+ \mathbf{U} \cdot \nabla \mathbf{U}  - G(t) \hat{\mathbf{x}} + \nabla P - \nu \Delta \mathbf{U} = - \overline{\mathbf{u}' \cdot \nabla \mathbf{u}'}~,
\label{eq:NSm}\\
 \partial_t\mathbf{u}'&+   \mathbf{U} \cdot \nabla \mathbf{u}' +
\mathbf{u}' \cdot \nabla \mathbf{U}  + \nabla \mathbf{p}' -  \nu \Delta  \mathbf{u}' 
= - \(  \mathbf{u}' \cdot \nabla \mathbf{u}' - \overline{\mathbf{u}' \cdot \nabla \mathbf{u}'} \,\)~, 
 \label{eq:NSp}\\
&\nabla \cdot \mathbf{U} = 0~,~~~\nabla \cdot \mathbf{u}' = 0~,\label{eq:NSdiv0}
\end{align}\label{eq:NSE0}\end{subequations}
where $\nu$ is the coefficient of kinematic viscosity. The $x,y,z$ components of $\U$ are $(U,V,W)$ and the corresponding components of $\u'$ are $(u',v',w')$. 
  The streak component of the
streamwise  mean flow is denoted with $U_s$ and defined as $U_s=U-[U]$. The $V$ and $W$  are the streamwise mean velocities of the roll vortices.
Streamwise mean perturbation Reynolds stress components are therefore written as e.g. $\overline{u'u'}$, $\overline{u'v'}$.
 
The RNL approximation is obtained by neglecting the perturbation-perturbation interaction terms in Eq.~\eqref{eq:NSp}. The RNL system is:
\begin{subequations}
\begin{align}
\partial_t\U&+ \U \cdot \nabla \U  - G(t) \hat{\mathbf{x}} + \nabla P - \nu\Delta \U = - \overline{\u' \cdot \nabla \u'}~,
\label{eq:QLm}\\
 \partial_t\u'&+   \U \cdot \nabla \u' +
\u' \cdot \nabla \U  + \nabla \mathbf{p}' -  \nu \Delta  \u' 
= 0~, 
 \label{eq:QLp}\\
&\nabla \cdot \mathbf{U} = 0~,~~~\nabla \cdot \mathbf{u}' = 0~.\label{eq:NSdiv0}
\end{align}\label{eq:QLE0}\end{subequations}
Equation~\eqref{eq:QLm} describes the dynamics of the streamwise mean flow, $\U$, 
which is driven by the divergence of the streamwise mean Reynolds stresses. These Reynolds stresses are obtained from Eq.~\eqref{eq:QLp} which incorporates the influence of the time dependent streamwise mean flow $\U(y,z,t)$ on  the streamwise varying perturbations $\u'$ but not the nonlinear interaction among the perturbations. Only the interaction of the perturbations directly on the streamwise mean  flow, $\U$, is retained in the rhs of Eq.~\eqref{eq:QLm}.
Remarkably, RNL self-sustains turbulence by incorporating this one essential nonlinear interaction, in the absence of which a self-sustained turbulent state cannot be established~\cite{Gayme-2010-thesis,Gayme-etal-2010}.

\begin{center}
\begin{table}[h]
\caption{\label{table:geometry}Simulation parameters. $[L_x,L_z]/h$ is the domain size in the streamwise, spanwise direction. $N_x$, $N_z$ are the number of Fourier components after dealiasing and $N_y$ is the number of Chebyshev components. $\Ret$ 
 is the Reynolds number of the simulation based on the friction velocity   and $[L_x^+$,$L_z^+]$ is the channel size in wall units.}
\centering
\begin{tabular}{@{}*{7}{c}}
\br
abbreviation &$[L_x,L_z]/h$&$N_x\times N_z\times N_y$& ${\Ret}$ &[$L_x^+$,$L_z^+]$\\
\mr
NS350   & $[\pi\;,\;\pi]$&$128 \times 255 \times 193 $  &357.1 & $[1122,1122]$\\
RNL350  &  $[\pi\;,\;\pi]$&$128 \times 255 \times 193$  & 353.5 & $[1111,1111]$\\
NS950   & $[\pi\;,\;\pi/2]$&$256\times 255\times 385$&939.9 & $[2953,1476]$\\
RNL950  &  $[\pi\;,\;\pi/2]$&$256\times  255\times 385$  & 882.4& $[2772,1386]$\\
\br
\end{tabular}
\end{table}
\end{center}

\section{Numerical approach and simulation parameters }
\label{sec:numerics}

The data were obtained from a DNS of Eqs.~\eqref{eq:NSE0} and from the RNL that is directly associated 
with the DNS.
Both the DNS and its directly associated RNL are integrated with no-slip boundary conditions in the wall-normal direction and periodic boundary conditions in the streamwise and spanwise directions. The dynamics were 
expressed in the form of evolution equations for the wall-normal vorticity and the Laplacian of the wall-normal velocity, with spatial discretization 
and Fourier dealiasing in the two wall-parallel directions and Chebychev polynomials in the wall-normal direction~\cite{Kim-etal-1987}. Time stepping was implemented using the third-order semi-implicit Runge-Kutta method. 

Quantities reported in outer units lengths are
scaled by the channel half-width, $h$,  and time by $h/\ut$ and the corresponding Reynolds number is
$\Ret= \ut h / \nu$ where $\ut= \sqrt{ \nu \left.\df U/\df y\right|_{\rm w}}$ ($\left.\df U/\df y\right|_{\rm w}$ is the shear at the wall) is the friction velocity.
Inner units  lengths are scaled by $h_{\tau} = \Ret^{-1} h$ and time by $\Ret^{-1} h/\ut$. 
Velocities scaled by  the friction velocity $\ut$ will be denoted with the superscript $+$, which indicates inner unit scaling. 

Parameters for the simulations presented are listed in Table~\ref{table:geometry}.


\section{Comparison of turbulence structure and dynamics diagnostics between NS and RNL}
\label{sec:results}

In this section we compare  turbulence diagnostics obtained from self-sustained turbulence in the RNL system, Eqs.~\eqref{eq:QLE0}, to diagnostics obtained from the parallel DNS of the NS, Eqs.~\eqref{eq:NSE0}. The geometry and resolution of the NS and RNL cases are given 
in Table \ref{table:geometry}.  Results are reported for Poiseuille turbulence at either 
$\Ret=350$ or $\Ret=950$.  The RNL simulations were initialized with an NS state and run until a steady state was established.
The RNL simulations produce self-sustained turbulence with the time mean estimated $\Ret$ values reported  in Table~\ref{table:geometry}, which are close to the  $\Ret$ values of the NS turbulent state.
Henceforth, the  RNL simulations will be identified with the $\Ret$ value of the corresponding  NS.
 
 \begin{figure}[t]
\centering
\includegraphics[width=.73\textwidth,trim = 8mm 13mm 15mm 5mm, clip]{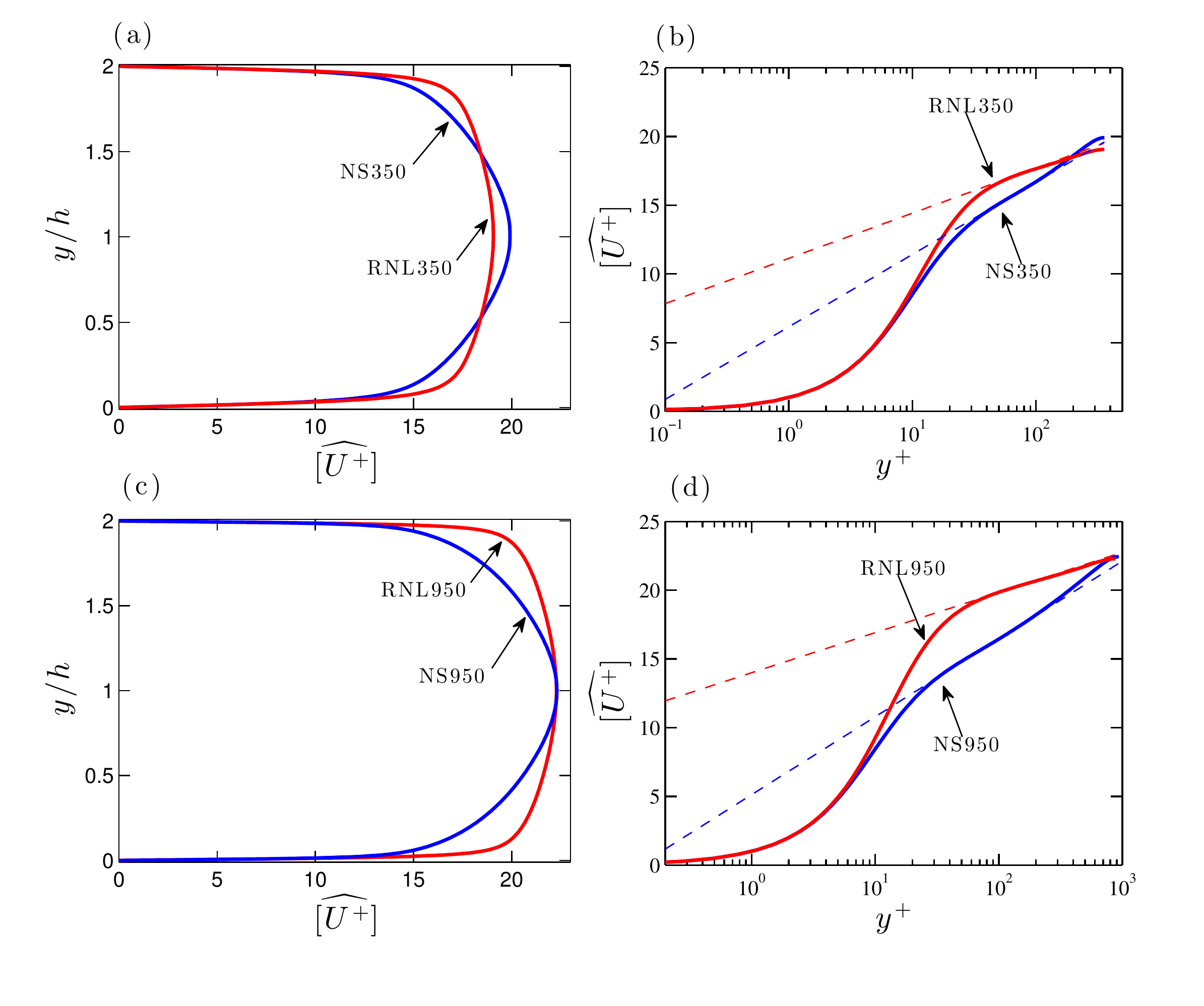}
\vspace{-.3cm}\caption{\label{fig:Uprofile350950}
Streamwise velocity $\widehat{\;[U^+(y)]\;}$ for the simulations in Table~\ref{table:geometry}. (a): NS350 and RNL350 simulations,  (c): NS950 and RNL950 simulations. In (b) and (d) are shown the corresponding  profiles  in wall units. The dashed lines  indicate the best fit  to  the law of the wall, $\widehat{\, [U^+]\,}=(1/\kappa) \log{(y^+)}+C$, with coefficients: NS350:~$\kappa = 0.44$, $C=6.1$, RNL350:~$\kappa=0.71$, $C=11.1$, NS950:~$\kappa=0.40$, $C=5.1$, RNL950:~$\kappa=0.77$, $C=14.0$.}
\end{figure}

\begin{figure}[h]
\vspace{-.1cm}
\begin{center}
\includegraphics[width=6.1in,trim = 11mm 2mm 10mm 2mm, clip]{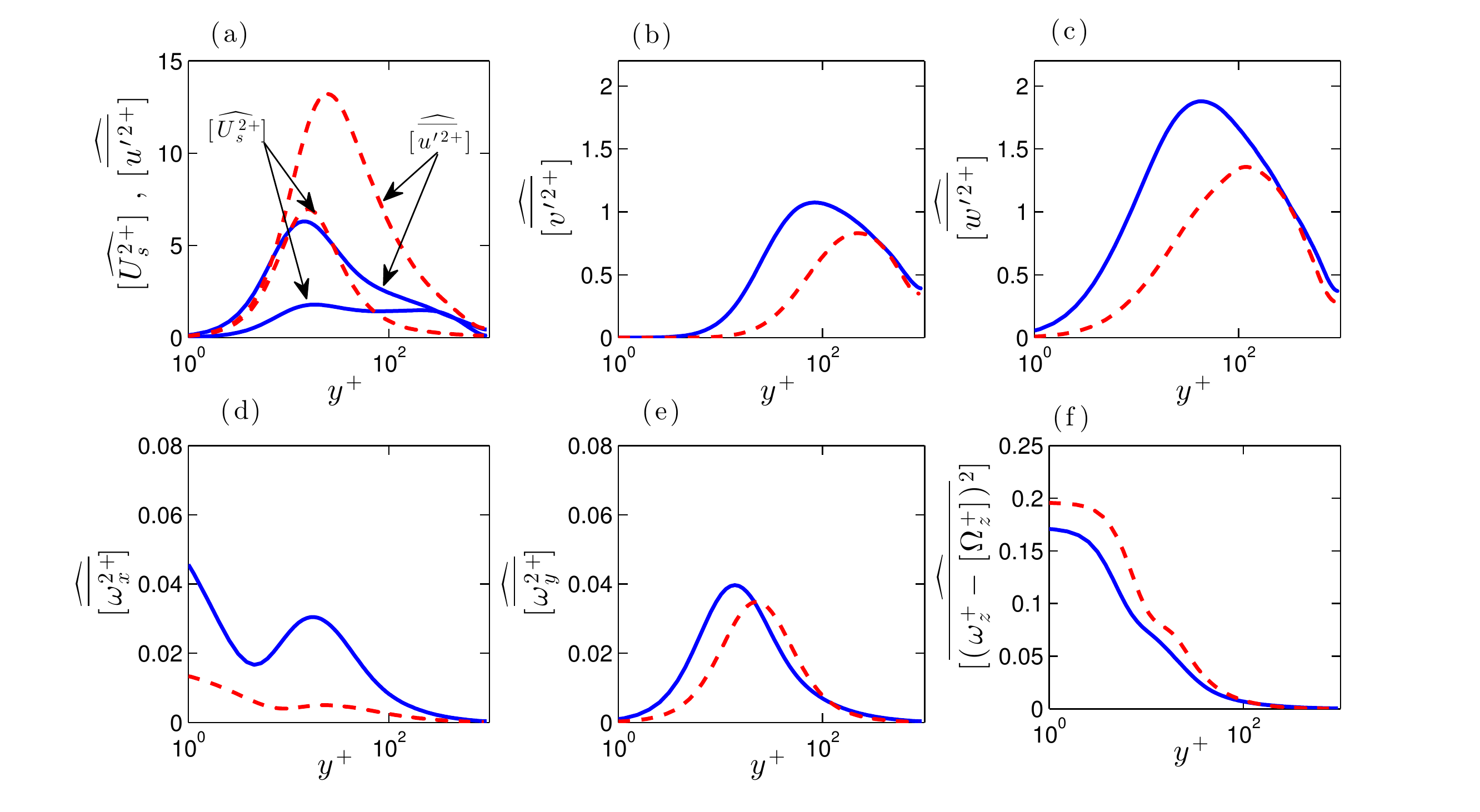}
\end{center}
\vspace{-.5cm}\caption{\label{fig:vel_om_prime} Comparison of velocity and vorticity fluctuations between NS950 and RNL950.  Shown are
 (a): $[\widehat{ U_s^{2+}}]$, $[ \widehat{\overline{u'^{2+}}} ]$, (b): $[\widehat{\overline{v'^{2+}}} ]$, (c): $[\widehat {\overline{w'^{2+}}} ]$, (d):  $[\widehat{\overline{\omega_x^{+ 2}}}]$, (e): $[\widehat{\overline{\omega_y^{+ 2}}}]$  and (f): $\widehat{\overline{ \[ (\omega_z^+-[\Omega^+_z])^2\]}}$, where $\Omega^+_z=-\partial U^+/\partial y^+$ for the NS950 (solid) and RNL950 (dashed).}
\end{figure}

The turbulent mean profiles for the NS and RNL simulations with $\Ret=350$ and  $\Ret=950$ are shown in outer variables in Fig.~\ref{fig:Uprofile350950}\hyperref[fig:Uprofile350950]{a},\hyperref[fig:Uprofile350950]{c} while Fig.~\ref{fig:Uprofile350950}\hyperref[fig:Uprofile350950]{b},\hyperref[fig:Uprofile350950]{d} shows the same data in  inner-wall units. 
Previous simulations in Couette turbulence~\cite{Thomas-etal-2013-sustain} at lower Reynolds numbers ($\Ret=65$) showed very small difference between the mean turbulent profile in NS and RNL. These simulations  at larger Reynolds numbers show significant differences in the mean turbulent profiles sustained by NS and RNL. This is especially pronounced in the outer regions where RNL  produces a mean turbulent profile with substantially smaller shear. Both profiles produce a log layer. However, the shear in these logarithmic regions
are different: the von K\'arm\'an constant of NS at $\Ret=950$  is $\kappa=0.4$ while for RNL it is $\kappa=0.77$. 
Formation of a log layer indicates that the underlying dynamics producing the log layer are retained in RNL.
Because RNL 
maintains essentially the same stress and  variance as NS in the log layer with a smaller shear, RNL
is in this sense more  efficient  than NS in transferring energy from the mean to the perturbations.

   
%

A comparison of perturbation statistics in RNL and NS is shown in Fig.~\ref{fig:vel_om_prime}
for $\Ret=950$.  The streamwise perturbation velocity fluctuations are significantly  more pronounced in RNL 
 and the magnitude of the streak in  RNL exceeds significantly the streak magnitude 
in NS in the inner wall region (cf. Fig.~\ref{fig:vel_om_prime}\hyperref[fig:vel_om_prime]{a}). 
In contrast, the wall-normal and  spanwise fluctuations in RNL are less pronounced than in NS (cf. Fig.~\ref{fig:vel_om_prime}\hyperref[fig:vel_om_prime]{b,c}) and similarly the 
streak fluctuations in the outer region are also less pronounced in RNL (cf. Fig.~\ref{fig:vel_om_prime}\hyperref[fig:vel_om_prime]{a}). 
The structure 
of the fluctuations of the vorticity components as a function of $y$ is shown in Fig.~\ref{fig:vel_om_prime}\hyperref[fig:vel_om_prime]{d,e,f}. The $\omega_z$ and $\omega_y$ fluctuations are similar in  NS and RNL.
The large $\omega_z$ fluctuations are associated with the shear of the streamwise velocity, while the $\omega_y$  fluctuations are associated with the streamwise streak structure. 
Only the $\omega_x$ fluctuations differ appreciably in amplitude between RNL and NS. This vorticity component is primarily associated with the fluctuations in the  streamwise roll circulations that are responsible for maintaining the streak which is central to sustaining the turbulence. 


%
\begin{figure}
\begin{center}
\includegraphics[width=4.6in,trim = 2mm 4mm 6mm 2mm, clip]{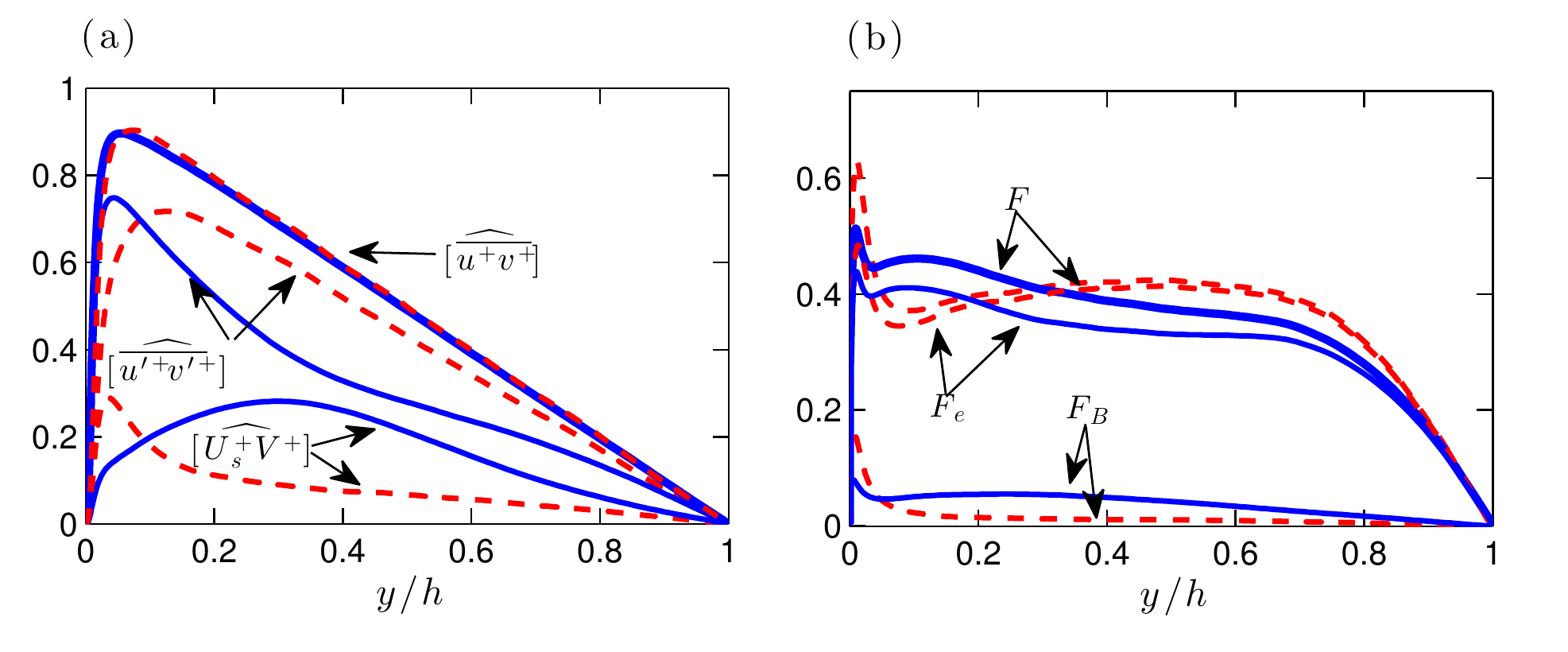}
\end{center}
\vspace{-5mm}\caption{\label{fig:uv_F}(a): The  Reynolds stress component, $\widehat{[\overline{uv}]}(y)$ in NS950 (solid) and  in RNL950 (dashed). 
Also shown are each of the terms, $\widehat{[U_sV]}$ and $\widehat{[\,\overline{u'v'}\,]}$ that sum to $\widehat{[\overline{uv}]}$. Although  
the NS and RNL values
of the total $\widehat{[\overline{uv}]}$ are almost identical, the contribution of  $\widehat{[U_sV]}$ and $\widehat{[\,\overline{u'v'}\,]}$  differ in NS and RNL. 
(b): Structure function, $F$, in NS950 (solid) and RNL950 (dashed). Shown are $F_B=-\widehat{[UV]} \big/ \sqrt{ \widehat{[U]^2} \widehat{[V]^2} }$,  $F_e=-\widehat{[\overline{u'v'}]} \big/ \sqrt{\widehat{[\overline{u'^2}]}\widehat{[\overline{v'^2}]}}$ and $F=-\widehat{[\overline{uv}]} \big/ \sqrt{\widehat{[\overline{u^2}]}\widehat{[\overline{v^2}]}}$.}
\vspace{2mm}
\end{figure}

\begin{figure}[h]
\begin{center}
\includegraphics[width= .52\textwidth,trim = 11mm 9mm 11mm 2mm, clip]{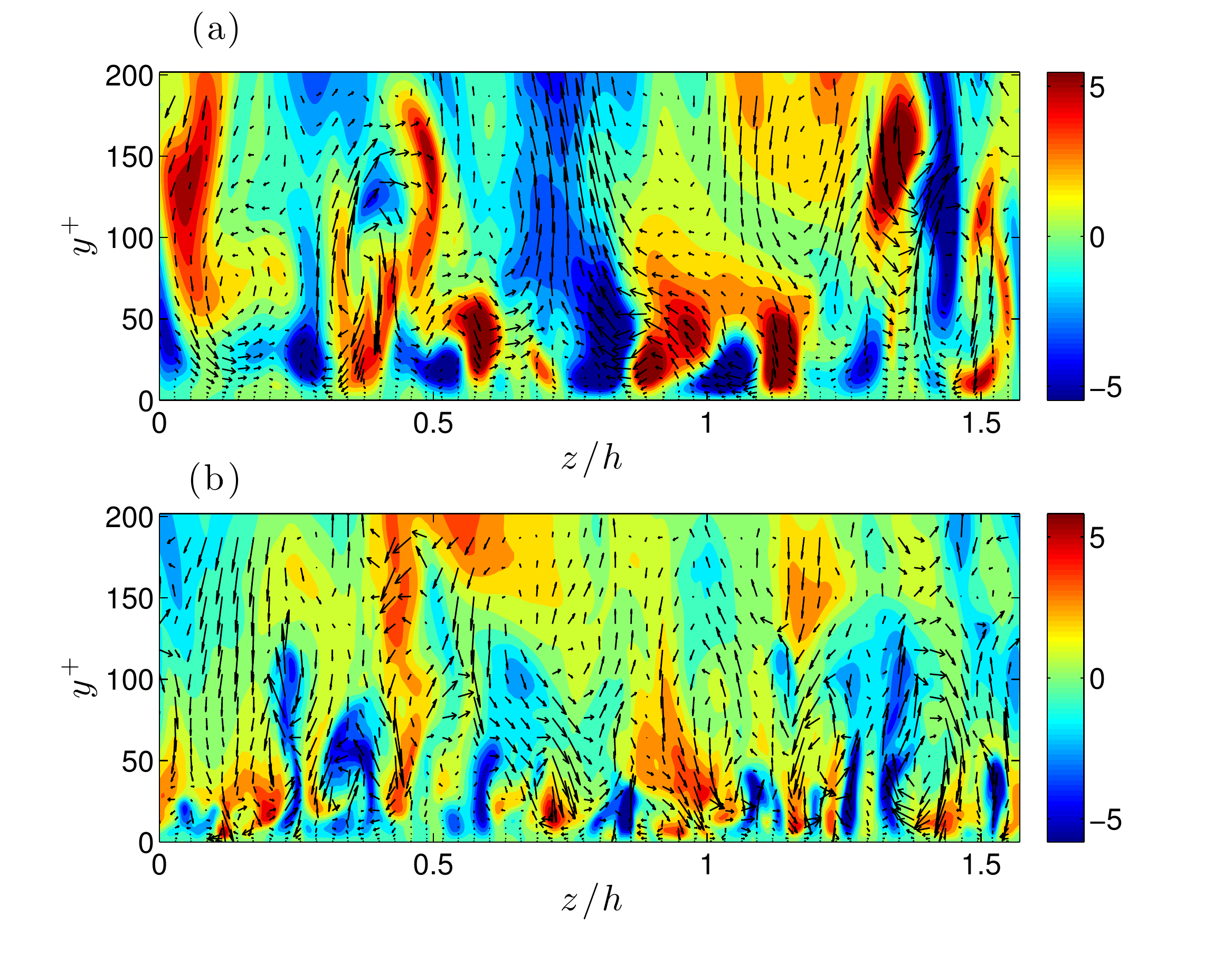}\end{center}
\vspace{-.8cm}\caption{\label{fig:YZpert}   Perturbation structure, $\u'^+$  in $(y,z)$ plane cross-section  for (a) RNL950 and (b) NS950 in the inner wall region, $0\le y^+\le200$. Both panels show contours of the $u'^+$ field, superimposed with components the $(v'^+,w'^+)$ velocities.}
\vspace{.5cm}
%
\begin{center}
\includegraphics[width= .98\textwidth,trim = 42mm 1mm 18mm 2mm, clip]{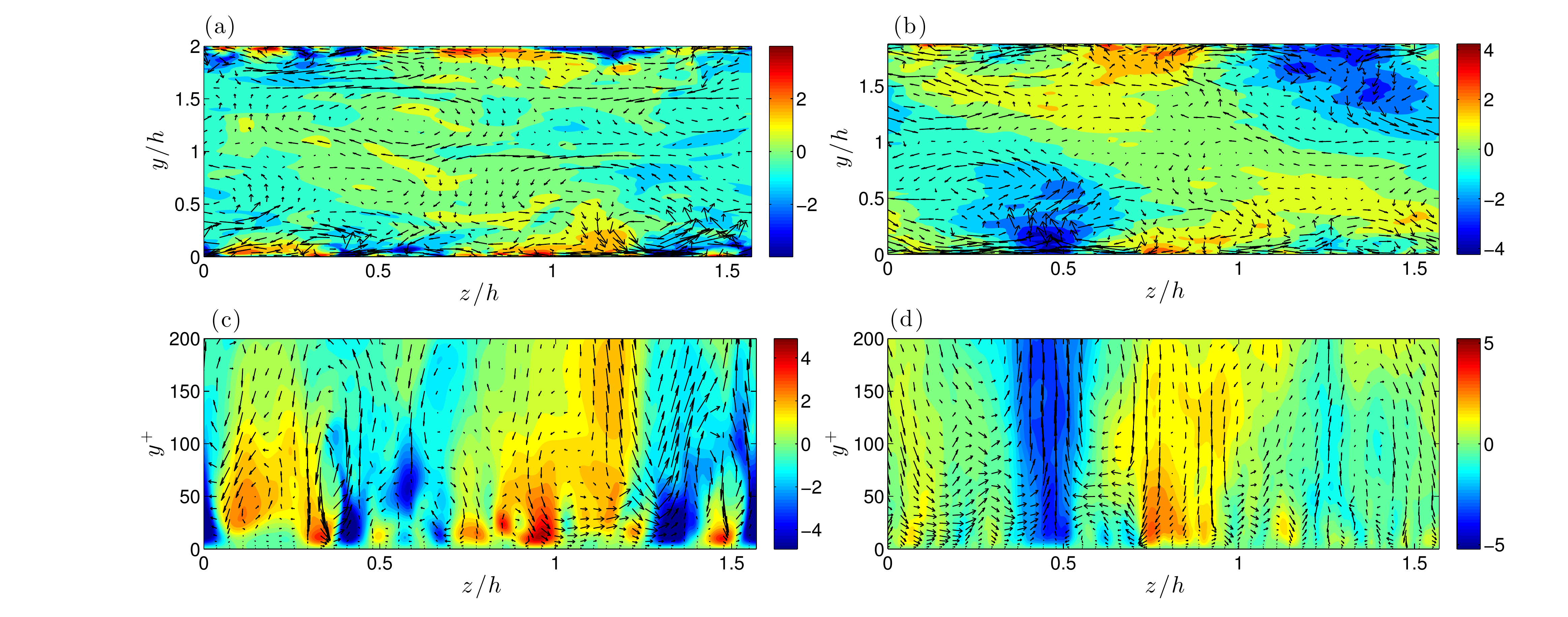}\end{center}
\vspace{-.8cm}   \caption{\label{fig:YZstreak}Instantaneous streamwise average flow velocity $\U^+$ (the $k_x=0$ component of the flow) shown as a $(y,z)$ plane cross-section for (a), (c) RNL950 and (b), (d) NS950. All panels show contours of the streak velocity, $U_s^+=U^+-[U^+]$, superimposed with the components of the $(V^+,W^+)$ velocities. The top panels show the whole channel while the bottom panels show the inner wall region, $0\le y^+\le200$.}
\vspace{.5cm}
\end{figure}

Despite this difference in the r.m.s. values of the velocity fluctuations,  both RNL and NS produce very similar $uv$ Reynolds stress. The increased amplitude of the velocity fluctuations in RNL is consistent with the fact that  RNL and NS produce nearly the same energy dissipation rate. The Reynolds stress  $\widehat{[\,\overline{uv}\,]}$ is the sum  of $\widehat{[U_s V]}$ and  $\widehat{[\,\overline{u'v'}\,]}$. 
Comparison of the wall-normal distribution of the time mean of these two components of the Reynolds stress is shown in Fig.~\ref{fig:uv_F}\hyperref[fig:uv_F]{a}.  Because the turbulence in NS and  RNL is sustained with essentially the same pressure gradient, the sum of these  Reynolds stresses  is the same linear function of $y$ outside the viscous layer. The Reynolds stress is dominated by the perturbation Reynolds stress $\widehat{[\,\overline{u'v'}\,]}$ in both simulations, with the RNL stress penetrating farther from the wall. This is consistent with the fact that the 
perturbation structure in RNL has larger scale.
This can be seen in a comparison of the NS and RNL perturbation structure  shown in Fig.~\ref{fig:YZpert}. 
Note that the Reynolds stress $\widehat{[U_sV]}$
associated with the streak and roll in the outer region of the NS simulation is larger than that in RNL. 
Further, the average correlation between the perturbation
$u'$ and $v'$ fields are  almost the same in both simulations while the correlation between the $U_s$ and $V$ 
in the RNL is much smaller than that in NS in the outer layer. 
This is seen in a plot of the structure function (cf. \cite{Flores-Jimenez-2006}) shown  in Fig.~\ref{fig:uv_F}\hyperref[fig:uv_F]{b}.

\begin{figure}[t]
  \centering
  \subfigure[{\bf \ NS\vspace{-.45cm}}\label{fig:dns_perspective}]{
   \includegraphics[width = .58\textwidth,trim = 43mm 7mm 43mm 8mm, clip]{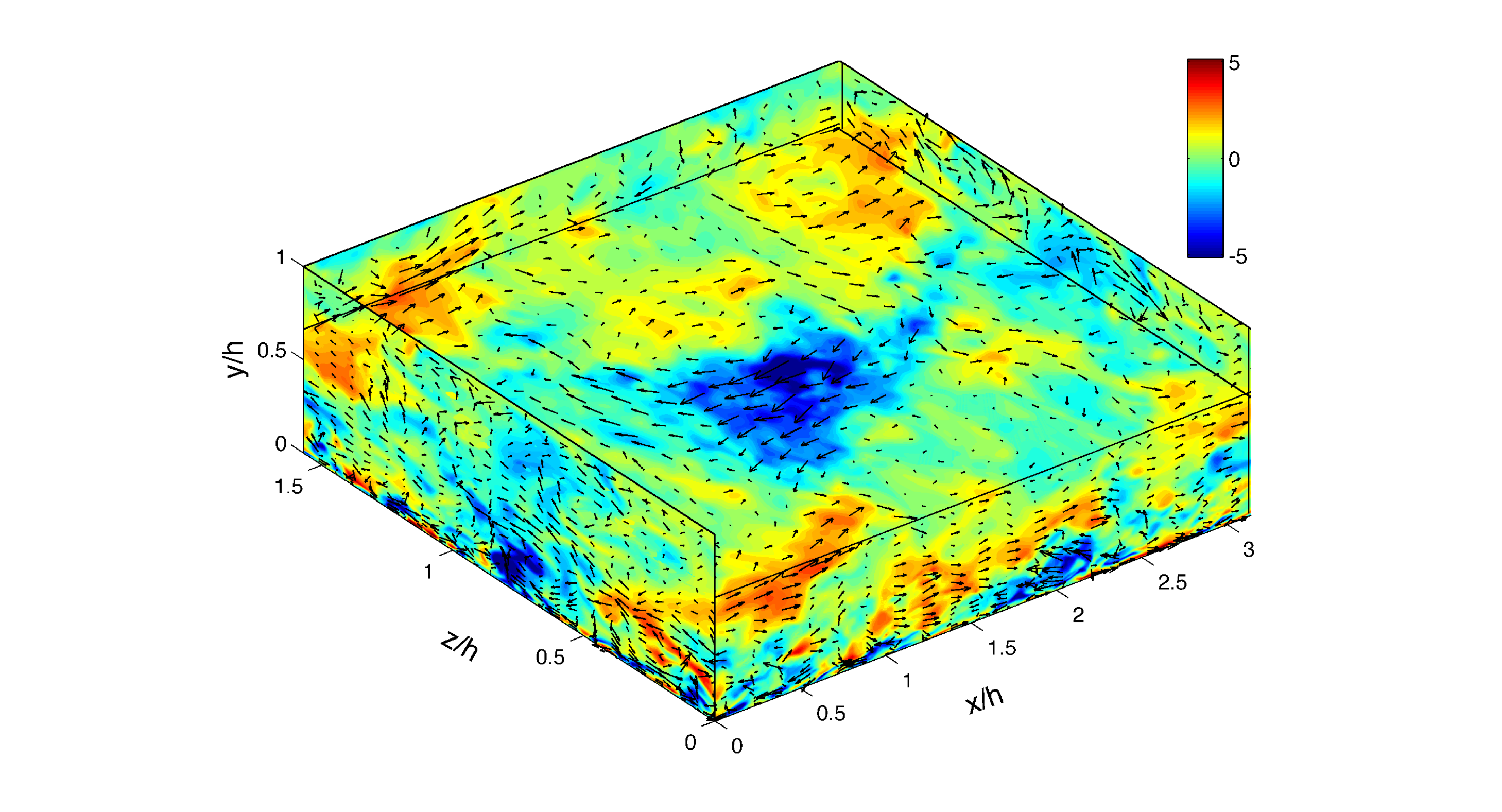}}
    \begin{minipage}{.38\textwidth}\vspace{-7.4cm}\caption{\label{fig:3Dstreak}3D perspective plots of the flow  at a single time for (a)  NS950, and (b)  RNL950 for lower half of the channel, $0\le y/h \le 1$. Both images show contours of the streak component plus streamwise perturbation, $U_s^++u'^+$. The central $x$-$z$ panel shows the flow at channel height, $y/h =0.65$.  The superimposed vectors represent the $(U_s^++u'^+,w^+)$ velocities for the $x$-$z$ panel, $(U_s^++u'^+,v^+)$  velocities for the $x$-$y$ panels and $(v^+,w^+)$ velocities for the $y$-$z$ panels. The parameters of the simulations are given in Table \ref{table:geometry}.}
\end{minipage}
  \subfigure[][{\bf \ RNL}\label{fig:QL_perspective}]{
   \includegraphics[width = .58\textwidth,trim = 43mm 7mm 43mm 8mm, clip]{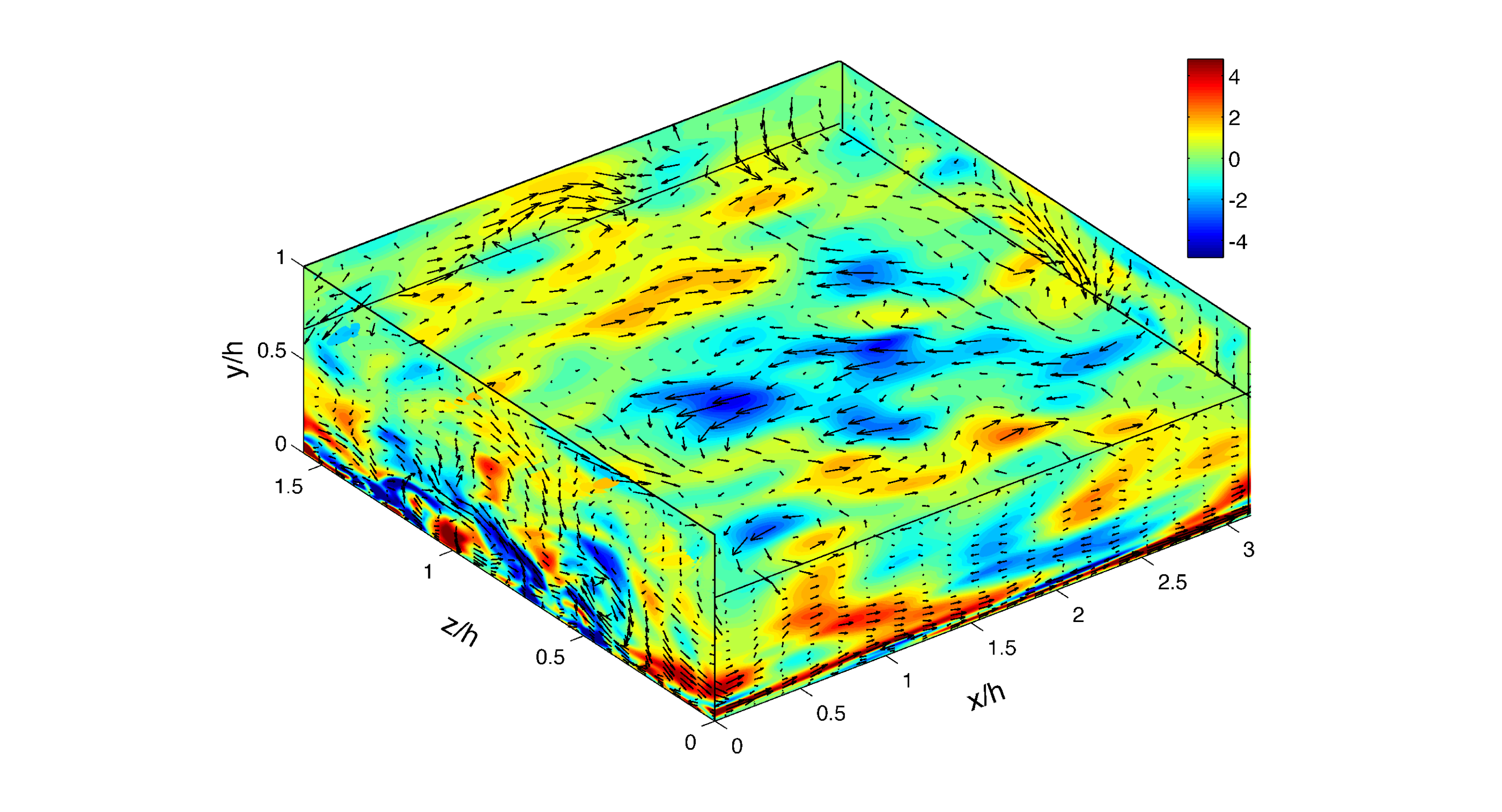}
    }
    \begin{minipage}[b]{14pc}\mbox{}
\end{minipage}
\end{figure}

Turning now to the NS950 and RNL950 simulations, a $(y,z)$ plane  snapshot of the streamwise mean flow component (corresponding to $k_x=0$ streamwise wavenumber) is shown in Fig.~\ref{fig:YZstreak}. Contours of  the  streamwise streak flow field, $U_s$, are shown together with vectors of the  streamwise mean $(V,W)$ field, which indicates the velocity components of the large-scale  roll structure. The presence of  organized streaks and associated  rolls is evident both in the inner-wall and in the outer-wall region. Figure~\ref{fig:3Dstreak} shows a three-dimensional perspective of the flow of the NS and RNL simulations, in which all of the  $k_x$ components of the velocity field are included. Note that in the RNL there is no visual evidence of the $k_x=0$ roll/streak structure which is required by the restrictions of RNL dynamics to be  the primary structure responsible for maintaining the RNL self-sustained turbulent state. Rather, the most energetic streamwise harmonic ($h k_x=2$ for this channel) is most prominent and dominates the perturbation structure.

\begin{figure}[t]
\begin{center}
\includegraphics[width=6in,trim = 0mm 20mm 0mm 0mm, clip]{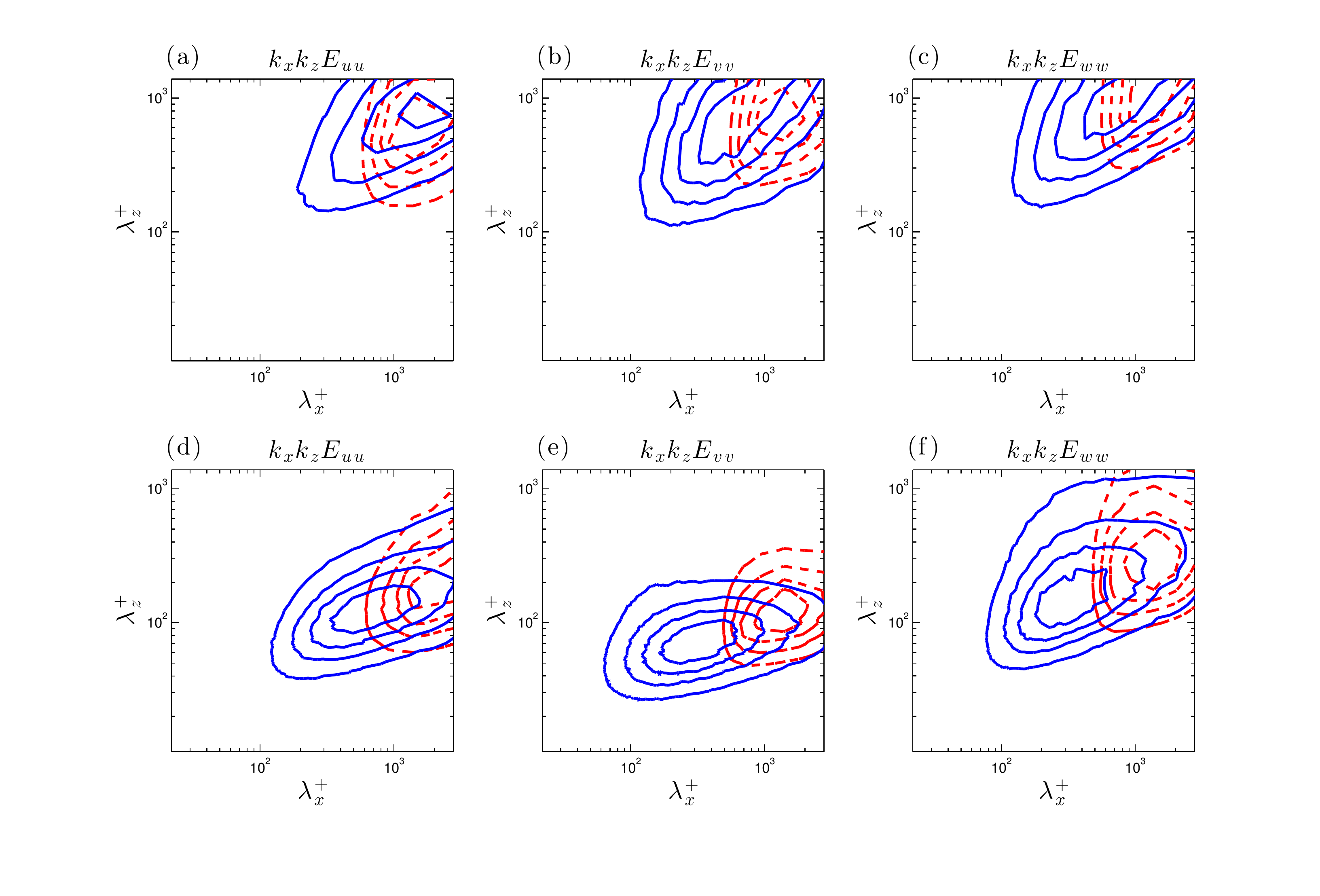}
\end{center}
\vspace{-.8cm}\caption{\label{fig:spectr_2d}
Contours of pre-multiplied power spectra $ k_x k_z E_{ff}(k_x,k_z)$ with $f=u,v,w$, as a function of $\la^+_x$ and $\la^+_z$ for NS950 (solid) and RNL950 (dashed). (a), (b) and (c) show the spectral energy densities at wall distance $y/h=0.65$ for the $u$, $v$ and $w$ respectively,  while panels (d), (e) and (f) show the corresponding spectral energy densities at $y^+=20$. Contours are (0.2,0.4,0.6,0.8) times the maximum value of the corresponding spectrum. The maximum $\la_x^+$ and $\la_y^+$ are  the lengths  $L_x^+$, $L_z^+$ of the periodic channel.}
\vspace{.5cm}
\end{figure}

\begin{figure}[t]
\begin{center}
\includegraphics[width=6.1in,trim = 10mm 8mm 10mm 5mm, clip]{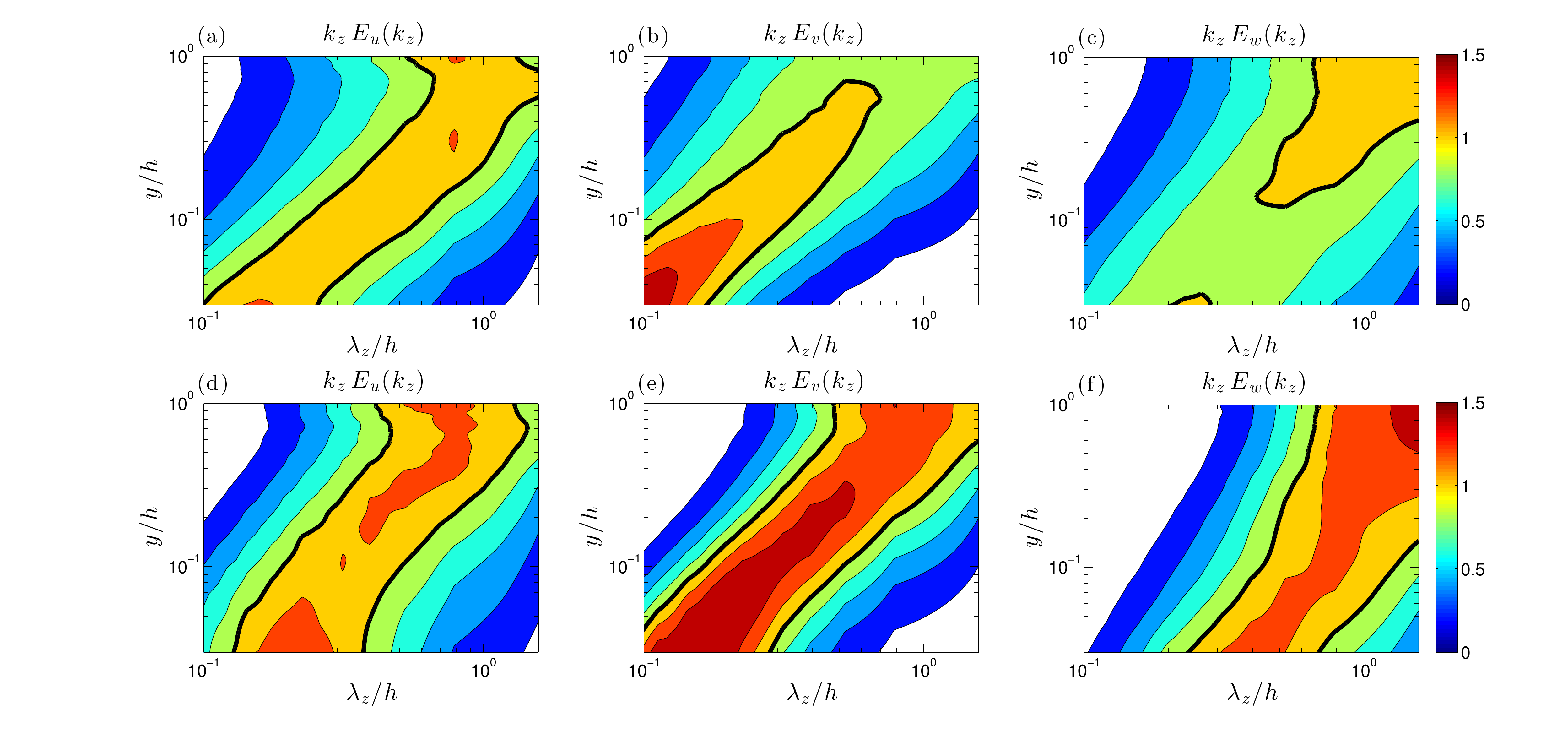}
\end{center}
\vspace{-.5cm}\caption{\label{fig:lambda_z}Normalized pre-multiplied spectral densities $k_z E_{f}(k_z) = k_z \sum_{k_x} E_{ff}(k_x,k_z)$, with $f=u,v,w$, as a function of spanwise wavelength, $\la_z/h$, and $y/h$. Spectral densities are normalized so that at each $y$ the total energy, $\sum_{k_z} E_{f}(k_z)$, is the same.  Shown are for NS950 (a): $k_z E_{u}(k_z)$, (b): $k_z E_{v}(k_z)$, (c): $k_z E_{w}(k_z)$ and for RNL950 (d): $k_z E_{u}(k_z)$, (e): $k_z E_{v}(k_z)$, (f): $k_z E_{w}(k_z)$. The isocontours are $0.2,0.4,\dots,1.4$ and the thick line marks the 1.0 isocontour.}
%
\begin{center}
\includegraphics[width=6.1in,trim = 10mm 8mm 10mm 5mm, clip]{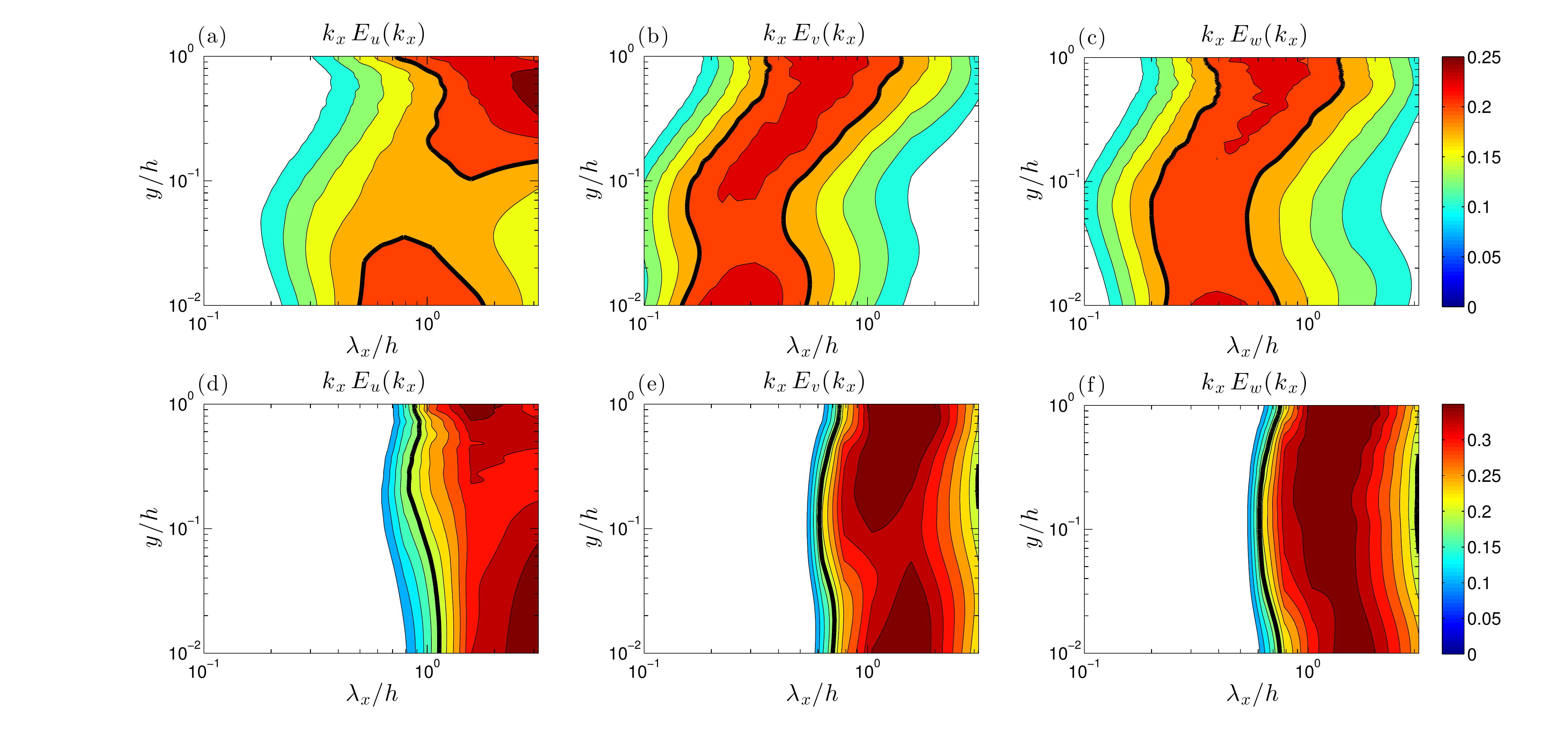}
\end{center}
\vspace{-.5cm}\caption{\label{fig:lambda_x}Normalized pre-multiplied spectral densities $k_x E_{f}(k_x) = k_x \sum_{k_z} E_{ff}(k_x,k_z)$, with $f=u,v,w$, as a function of streamwise wavelength, $\la_x/h$, and $y/h$. Spectral densities are normalized so that at each $y$ the total energy, $\sum_{k_x} E_{f}(k_x)$, is the same. Shown are for NS950 (a): $k_x E_{u}(k_x)$, (b): $k_x E_{v}(k_x)$, (c): $k_x E_{w}(k_x)$ and for RNL950 (d): $k_x E_{u}(k_x)$, (e): $k_x E_{v}(k_x)$, (f): $k_x E_{w}(k_x)$. The isocontours are $0.1,0.125,\dots,0.35$ and the thick line marks the 0.2 isocontour.}
\end{figure}

A comparison of the spectral energy densities  of the velocity fields as a function of   streamwise and spanwise wavenumber, $(k_x,k_z)$,  provides an alternative view of the turbulent structure.  The premultiplied spectral energy densities of each of the three contributions to the kinetic energy, $E_{uu}$, $E_{vv}$ and $E_{ww}$ at channel
heights $y^+=20$,  representative of the inner-wall region, and $y/h =0.65$, representative of the outer-wall region, are shown in Fig.~\ref{fig:spectr_2d}.
Near the inner-wall,  RNL produces  spanwise streak spacing and rolls similar to those in NS. The tendency of RNL to favor longer
structure is also evident in these figures.  The spectra for the outer region indicate similar large scale structure and good agreement in
the spanwise spacing between RNL and NS. This figure  establishes the presence of 
large-scale structure  in the outer region in both RNL and NS.  
The corresponding contour plots of normalized premultiplied one-dimensional spectral  energy densities as a function of spanwise wavelength and wall-normal distance, shown in Fig.~\ref{fig:lambda_z}\hyperref[fig:lambda_z]{a,c,d,f} 
 accord  with the spectra of Toh~\&~Itano~\cite{Toh-Itano-2005} and demonstrate
 again the similarity in structure in NS and RNL and specifically the presence of  large structures in the outer flow.   
It has been noted that in NS while  the dominant large scale structures scale linearly with distance from the wall in the inner-wall region, 
in the outer regions structures having the largest possible streamwise scale dominate the flow variance at high 
Reynolds number~\cite{Jimenez-1998,Jimenez-Hoyas-2008}. This linear scaling near the wall  can be seen in  
Fig.~\ref{fig:lambda_z} where the pre-multiplied spectral densities as a function of the distance $y$ and of the spanwise wavelength, $k_z$, are shown, as in~\cite{Jimenez-1998,Jimenez-Hoyas-2008}, for both NL and RNL. In both simulations the spanwise wavelength associated with the spectral density maxima increases linearly with wall distance and this
linear dependence  is intercepted at $y/h\approx 0.5$ (or $y^+\approx 450$).  Beyond $y/h\approx 0.5$ 
structures assume the widest wavelength allowed in the channel, suggesting that simulations  must be performed in longer 
boxes in future work (cf. discussion in Jim\'enez \& Hoyas~\cite{Jimenez-Hoyas-2008} and Flores \& Jim\'enez~\cite{Flores-Jimenez-2010}).
Corresponding contour plots of spectral energy density as a function of streamwise wavelength and wall-normal distance  are shown in  Fig.~\ref{fig:lambda_x}. These plots show that the perturbation variance  in the inner wall and outer wall region is concentrated in a limited set of  streamwise components which is apparent in Fig.~\ref{fig:3Dstreak}.
The  restriction of the streamwise structure is particularly pronounced in the case of  RNL in which
the  outer layer variance peaks at $h k_x=4$,  which scale  the wall-normal velocity inherits.
 Note that the maximum wavelength in these graphs is equal to the streamwise length of the box and not to the infinite wavelength associated with the energy 
of the streak/roll structure, which as we will see is the most energetic structure in the DNS but not in the RNL.   

%

%

\section{The RNL system as a minimal turbulence model}

\begin{figure}[b]
\begin{center}
\includegraphics[width=5.5in,trim = 0mm 6mm 0mm 1mm, clip]{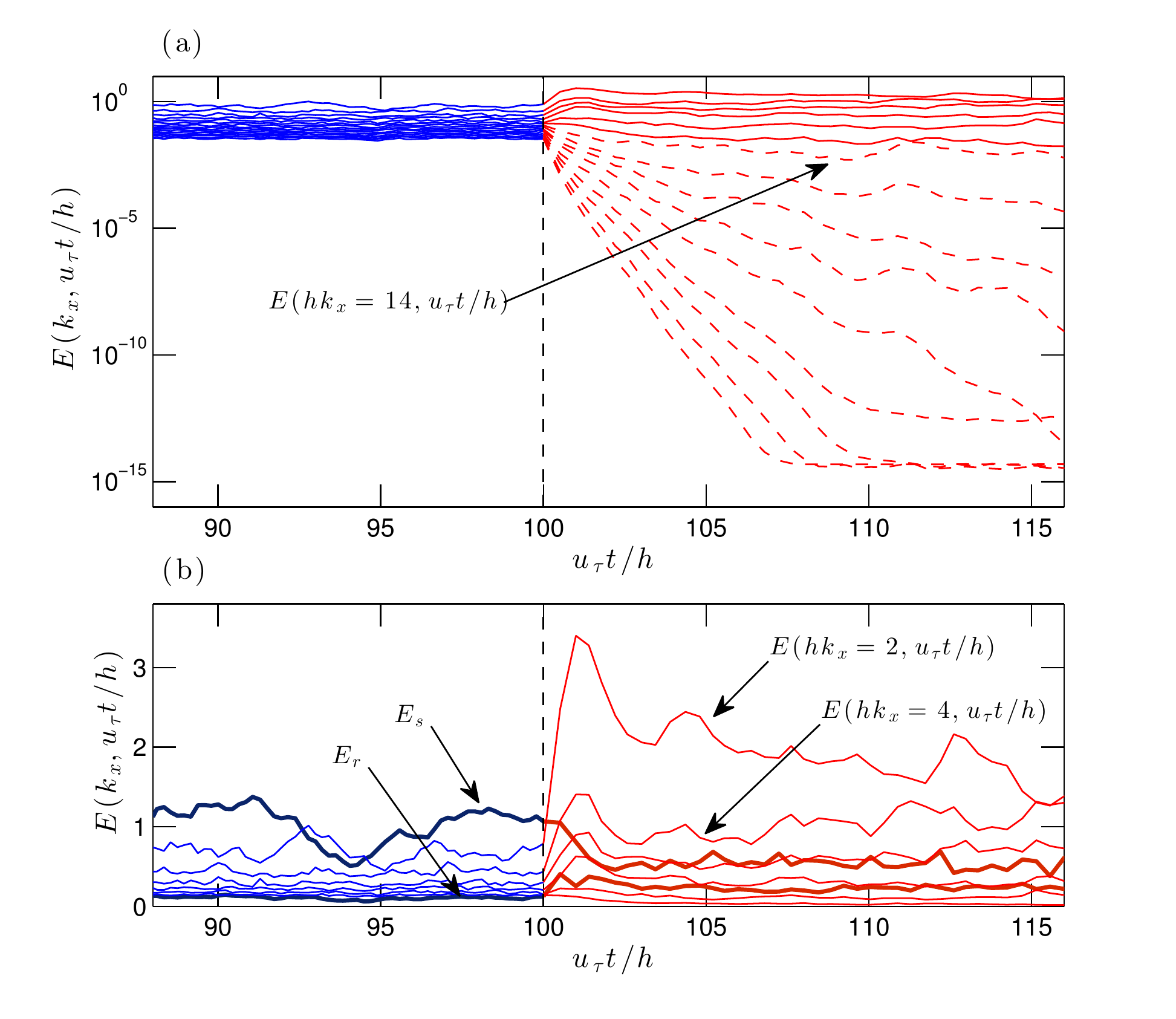}
\end{center}
\vspace{-.5cm}\caption{\label{fig:Enkx_DNSQL950}
An  NS950 simulation up to $\ut t / h=100$ (indicated with the vertical line) is continued subsequently under RNL dynamics. 
Shown are  (a) Energy of the  first 15 streamwise varying  Fourier components ($h k_x=2,4,\dots,30$). The energy of the Fourier components decreases monotonically with wavenumber. Decaying   Fourier components are indicated with dashed lines. After the transition to RNL  dynamics all components with $h k_x \ge 14$  decay ($h k_x=14$  decays, although it is not shown in this figure).  Asymptotically the dynamics of the RNL950  turbulence is maintained by interaction between   the set of surviving $h k_x=2,4,\dots,12$  Fourier components and the mean flow ($k_x=0$).  (b)  
Detailed view showing the energy of the mean and surviving perturbation components during the transition from NS to RNL dynamics, in which the total energy increased by $10\%$.
For the $k_x=0$ shown are:  the streak energy, $E_s=(hL_z)^{-1}\int \df y\,\df z\, U_s^{+2}/2$, and roll energy, $E_r=(h L_z)^{-1}\int \df y\,\df z\, (V^{+2}+W^{+2})/2$. The energy  of the $h k_x=2,4,6,8$ components  increases rapidly  during the adjustment after transition to RNL dynamics.  Note that the total energy in the perturbation $k_x\ne 0$ components decreases from $0.91$ in the NS950 ($0.56$ being in the components that survive in the RNL)  to $0.78$   in RNL950. Also the roll/streak energy  decreases from  $1.1$ in NS950 to $0.8$ in RNL950, while the  energy of the $k_x=k_z=0$ component increases from $397$  to $448$. }
\end{figure}

RNL turbulence has the property that when initiated with full NS turbulence
it spontaneously transitions to a self-sustaining turbulent state
supported by a severely restricted set of streamwise Fourier components. 
This property is consistent with the structure of the RNL system, which retains only the interaction between the 
$k_x=0$ and the $k_x \ne 0$ components, from which it follows that the only energy source for 
maintaining a perturbation is its interaction with the mean flow~\cite{Farrell-Ioannou-2012}.  It is remarkable that at these fairly high Reynolds numbers only  a small set of streamwise harmonics are maintained by this interaction with the mean flow. Moreover, this is the fundamental interaction maintaining RNL turbulence and by implication, given the similarity in the structure and dynamics between them, of NS turbulence as well.  Even if the RNL dynamics is initialized with an NS flow state with energy in all $N_x$ components, the RNL turbulence eventually reduces to involve only the $n_x\ll N_x$  Fourier components with 
wavenumber $h k_x =(2\pi/L_x)\times( 0,1,\dots,(n_x-1))$.  We view this transition of  NS turbulence to RNL turbulence 
as a process of distillation by which a small set of structures 
maintaining the turbulent state is identified, a result that was previously obtained in the case of self-sustained Couette turbulence at $\Re=400$ and $\Re=1000$~\cite{Farrell-Ioannou-2012,Farrell-Ioannou-2012-CTRv2}.  In this previous work, 
a minimal channel RNL simulation at $\Re=400$ was shown to self-sustain turbulence 
by the interaction of only two $k_x$ components: $k_x=0$ and the first harmonic in the channel. The analogous distillation process for the $\Ret=950$ simulation is shown in Fig.~\ref{fig:Enkx_DNSQL950}\hyperref[fig:Enkx_DNSQL950]{a}. The time evolution of the energy of the first $15$ streamwise varying 
Fourier components in an NS simulation is shown in the left part of the figure ($\ut t/h<100$) while in the right part is shown the subsequent evolution of these components when at the indicated time the perturbation-perturbation interactions are suppressed so that the turbulence evolves under RNL dynamics. 
It is evident from Fig.~\ref{fig:Enkx_DNSQL950}\hyperref[fig:Enkx_DNSQL950]{a} that RNL950 turbulence in a channel with $L_x=\pi h$ retains only  the six  Fourier components
$h k_x=2,4,\dots,12$ out of the $N_x= 127$ streamwise varying components that are present in the NS simulation. All components with $h k_x >12$  decay exponentially.  As a result, a transition  occurs in which a reduced complexity dynamics maintaining turbulence arises, which self-sustains turbulence despite this greatly restricted support in streamwise wavenumber space. That RNL maintains a turbulent state similar to that of NS with nearly the same $\Ret$  ($\Ret=882$ vs. $\Ret=940$)  implies that these systems have approximately the same energy production and dissipation and that the $n_x$ components retained in RNL  assume the burden of accounting for
this energy production and  dissipation. Specifically, the components in NS  that are not retained in RNL are responsible for approximately  $1/3$ of the total energy dissipation, which implies that the components that are retained in RNL must increase their dissipation by that much. 

\section{Streak structure dynamics in NS and RNL}

Large scale roll/streak structures are prominent  in the inner layer as well as in the outer layer both in NS and in RNL. The dynamics of this structure  can be diagnosed using the time evolution of the energy of each $k_x$ component during the transition from NS to RNL, shown in  Fig.~\ref{fig:Enkx_DNSQL950}\hyperref[fig:Enkx_DNSQL950]{b}. 
It can be seen that in NS the energy associated with the streamwise mean structure with $k_x=0$ and $k_z\ne 0$ is dominant among the structures that deviate from the mean flow, $[U]$. In the inner layer the interaction of roll/streak structures with the $k_x\ne0$ perturbation field  maintains turbulence through an SSP~\citep{Hamilton-etal-1995, Jimenez-Pinelli-1999,Farrell-Ioannou-2012}. The RNL system provides an especially simple manifestation of this SSP as its dynamics comprise only interaction between the mean ($k_x=0$) and perturbation ($k_x \ne 0$) components. The fact that RNL self-sustains a counterpart
turbulent state provides strong evidence that the RNL SSP captures the essential dynamics of turbulence in the inner wall region.

The structure of the RNL system compels the interpretation that the time dependence of the SSP 
cycle, which might appear at first to consist of a concatenation of random and essentially unrelated events, is instead an intricate  interaction dynamics among streaks, rolls and perturbations that produces the $\U(y,z,t)$ which, when introduced in Eq.~\eqref{eq:QLp}, results in generation of an evolving perturbation Lyapunov structure with exactly zero Lyapunov exponent. 
 S3T  identifies this exquisitely contrived SSP cycle which  
 comprises  the generation of the streak  through lift-up by the rolls, the maintenance of the rolls by torques induced by the perturbations which themselves are maintained by an essentially time-dependent parametric non-normal interaction with  the streak (rather than e.g.~inflectional instability of the streak structure)~\citep{Farrell-Ioannou-2012}.

\begin{figure}[t]
\centering
\includegraphics[width= .56\textwidth,trim = 12mm 2mm 14mm 0mm, clip]{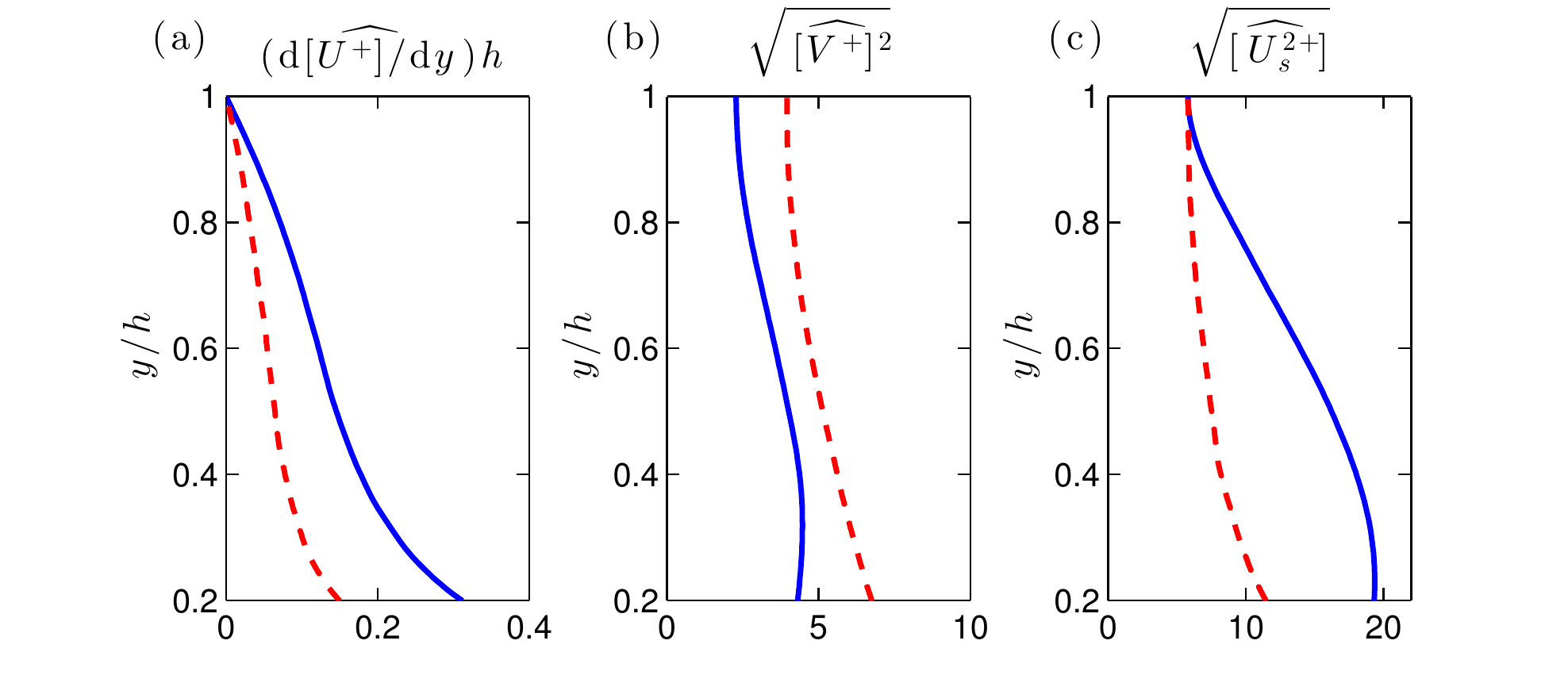}\hspace{.02\textwidth}%
\begin{minipage}{.41\textwidth}\vspace{-4cm}\caption{\label{fig:UyVrmsUsrms}
Comparison of the turbu\-lent mean shear, $(\widehat{\df [ U^+]/\df y})h$ (panel (a)), the r.m.s. of $[V^+]$ (panel (b)) and the r.m.s. of the streak velocity, $U_s^+$, (panel (c)) for  NS950 (solid) and RNL950 (dashed)  in the outer layer, $0.2\le y/h\le 1$.}
\end{minipage}
\end{figure}

Vanishing of the Lyapunov exponent associated with the SSP  is indicative  of a feedback control process acting between the streaks and the perturbations by which the parametric instability of the perturbations on the time dependent streak is reduced so that it asymptotically maintains a zero Lyapunov exponent. Examination of the transition from the NS to the RNL, shown by the simulation diagnostics in Fig.~\ref{fig:Enkx_DNSQL950}\hyperref[fig:Enkx_DNSQL950]{b}, reveals the action of this controller. When in Eq.~\eqref{eq:NSp} the interaction among the perturbations is switched off,  so that the simulation is governed by RNL dynamics, we observe a sudden increase of the energy of the surviving  $k_x \ne 0$ components, cf. the rapid increase of the energy in the $h k_x=2,4,6,8$ components. 
Increase of the energy of these components is expected because the dissipative effect of the  perturbation-perturbation nonlinearity that acts on   these  components is removed in RNL. As these modes grow
the SSP cycle quickly adjusts to a new turbulent equilibrium state characterized by reduced streak amplitude and 
increased energy in the largest streamwise scales. 
This SSP cycle is more efficient in the sense that with smaller mean shear a self-sustained turbulence with approximately the same $\Ret$ as that in NS is maintained.
This turbulent state is dominated by high amplitude fluctuations of the $h k_x =2,4,6,8$ components, as well as in the components associated with the wall-normal and spanwise  direction. 
This can be seen in a comparison of the NS and RNL perturbation structure (the velocity field corresponding to $k_x \ne 0$) shown in Fig.~\ref{fig:YZpert}. The perturbations in RNL simultaneously reduce the shear of the mean flow and maintain a reduced amplitude streak in the outer layer.
A comparison of the shear, of the r.m.s. $V$ velocity, and of the r.m.s. streak velocity, $U_s$,  in the outer layer is shown as a function of $y$ in Fig.~\ref{fig:UyVrmsUsrms}, from which it can be seen that the reduction of the  amplitude of the streak in RNL is equal to the reduction in the mean flow shear. It is important to note that these dependencies are integral to the SSP cycle and specifically of its feedback control which determines the statistical steady state and must be understood in the context of the cycle.

%

In the discussion above we have assumed that the presence of roll and streak structure in the outer layer in RNL indicates the existence of an SSP cycle there as well, and by implication also in NS.  In order to show this  consider the momentum equation for the streamwise streak:
\begin{align}
\partial_t U_{ s}  &=\underbrace{  -\(V \,\partial_y U -\[V \,\partial_yU\]\bit\) +\( W \,\partial_z U-\[ W \,\partial_z U\]\bit\)}_{\textrm{A}} -\nonumber\\
&\qquad\qquad \underbrace{-\(\overline{ v'  \,\partial_y u'-\[v'  \,\partial_y u'\]}\)- \(\overline{ w'  \,\partial_z u'-\[w'  \,\partial_z u'\]}\)}_{\textrm{B}}+ \,\nu\, \Del U_{ s}~.\label{eq:Us}
\end{align}
Term~A in Eq.~\eqref{eq:Us} is the contribution to the streak acceleration  by   the `lift-up' mechanism  and the `push-over' mechanism,  which represent transfer to streak momentum  by  the mean wall-normal and spanwise velocities respectively. Term~B in Eq.~\eqref{eq:Us} is the contribution to the streak momentum by the perturbation  Reynolds stress  divergence (structures with $k_x \ne 0$).


\begin{figure}
\centering
\includegraphics[width= .85\textwidth,trim = 18mm 10mm 22mm 8mm, clip]{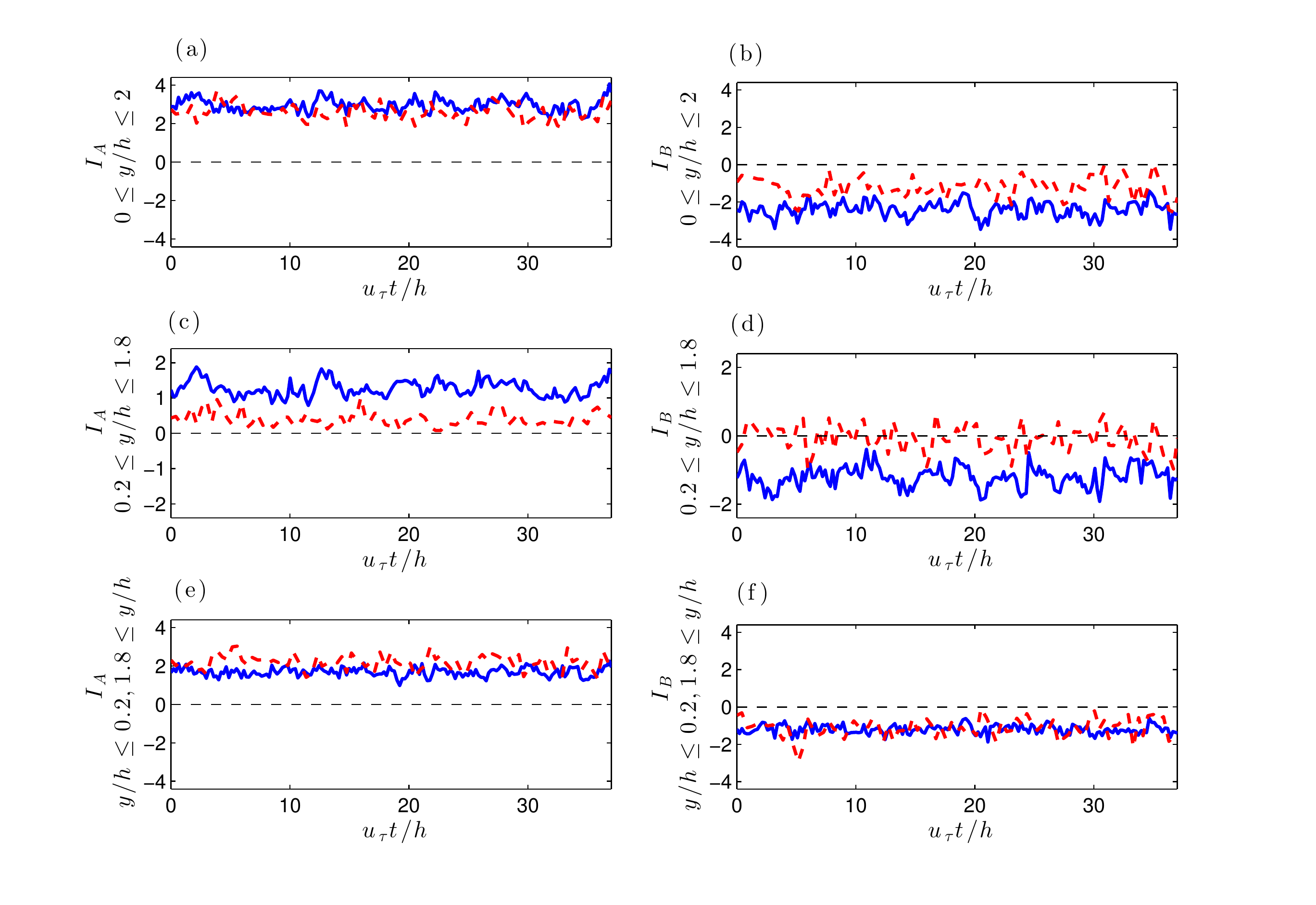}
\caption{\label{fig:Usdot_time} (a):  
$I_A(t)$ over the whole channel  with  time mean value $2.95$ for NS950 and $2.59$ for RNL950. (c): $I_A(t)$ over the outer region, $0.2\le y/h\le 1.8$, with  time mean value $1.97$ for NS950 and $0.4$ for RNL950.
(e) $I_A(t)$ over the inner region, $0\le y/h\le 0.2~1.8\le y/h\le 2$, with  time mean value $1.68$ for NS950 and $2.19$ for RNL950.
(b): $I_B(t)$ over the whole channel  with  time mean value $-2.4$ for NS950 and $-1.21$ for RNL950.
(d): $I_B(t)$ over the outer region, $0.2\le y/h\le 1.8$, with  time mean value $-1.2$ for NS950 and $-0.08$ for RNL950.
(f) $I_B(t)$ over the inner region, $0\le y/h\le 0.2~1.8\le y/h\le 2$, with  time mean value $-1.2$ for NS950 and $-1.13$ for RNL950.}
\end{figure}
\begin{figure}[t]
\centering
\includegraphics[width= 5.3in,trim = 6mm 8mm 1mm 1mm, clip]{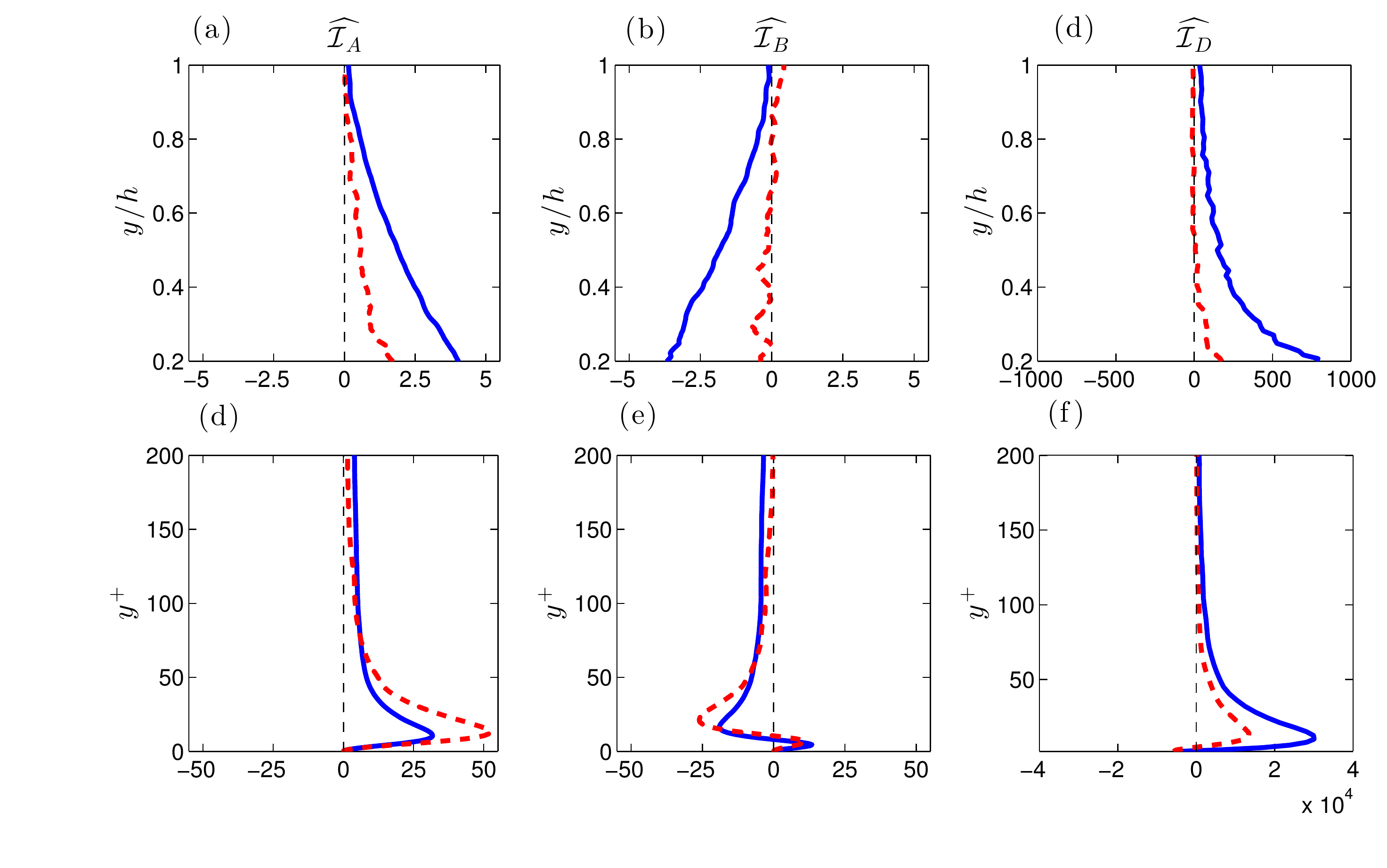}
\vspace{-.4cm}
\caption{\label{fig:UsdotyOMxdoty} (a) Contribution  to streak acceleration from the lift up mechanism $\widehat{\mathcal{I}_A}$. (b) Contribution to streak acceleration from the perturbation Reynolds stress divergence  $\widehat{\mathcal{I}_B}$.
(c) contribution to streamwise mean vorticity generation from perturbation Reynolds stress induced torques  $\widehat{\mathcal{I}_D}$ (cf.~section~\ref{sec:roll}). Results from NS950 (solid) and RNL950 (dashed). 
Upper panels show structure in  the outer layer, $0.2\le y/h\le 1$, lower panels show the structure in the inner layer, $0\le y^+\le 200$.}
\end{figure}

In order to identify the mechanism of streak maintenance we examine whether terms~A~and~B  accelerate or decelerate  the streaks by evaluating from the simulation the contribution of the terms 
$I_A (t) = h^{-1}\int_R\df y\;\mathcal{I}_{A}(y,t)$ and $I_B (t) = h^{-1}\int_R\df y\;\mathcal{I}_{B}(y,t)$,  where $\;\mathcal{I}_{A}(y,t) = h \ut^{-2}\,L_z^{-1}\int \df z\;\sgn(U_s) \times \textrm{(Term~A)}$ and $\;\mathcal{I}_{B}(y,t) = h \ut^{-2}\,L_z^{-1}\int \df z\;\sgn(U_s) \times \textrm{(Term~B)}$, to streak mainte\-nance as a function of time and region $R$ of the flow
(similar results are obtained by multiplying  with $U_s$ to obtain an energy rather than a momentum budget for $U_s$). The results of this calculation are shown in Fig.~\ref{fig:Usdot_time} for $R$ extending over the
the inner region, the outer region and the whole channel.  In the inner and outer wall regions for both NS and RNL the streak structure is supported primarily by  the lift-up mechanism while the Reynolds stress divergences oppose the streak.  While the magnitude of the acceleration by the lift-up and the deceleration by the Reynolds stress divergence are nearly the same in both NS and RNL in the inner region, in the outer region the 
acceleration by the lift-up in the RNL is about half of that in the NS  due to the smaller mean flow shear in RNL. The wall-normal structure of the time mean
$\widehat{\mathcal{I}_A}$ and  $\widehat{\mathcal{I}_B}$ are shown 
in Fig.~\ref{fig:UsdotyOMxdoty}\hyperref[fig:UsdotyOMxdoty]{a,b}. We conclude that in NS and RNL the only positive contributions to the outer layer streaks are  induced by  the lift-up from  the roll circulation, despite the small shear in this region. We next consider the dynamics maintaining the roll circulation.

%
%
%


\section{Roll dynamics: maintenance of mean streamwise vorticity in NS and RNL\label{sec:roll}}

We have established that the roll circulation is not only responsible for streak maintenance in the inner layer but also in the outer layer. We now examine the mechanism of  roll maintenance using  streamwise averaged vorticity, $\Omega_x=\partial_yW-\partial_zV$, in the outer layer as a diagnostic.
In order for roll circulation to be maintained against dissipation there must be a continuous generation of $\Omega_x$. There are two possibilities for the maintenance of $\Omega_x$ in the outer layer: either $\Omega_x$ is generated locally in the outer layer, or it is advected  from the near wall region.

\begin{figure}[b]
\centering
\hspace{.02\textwidth}\includegraphics[width= .54\textwidth,trim = 7mm 11mm 19mm 4mm, clip]{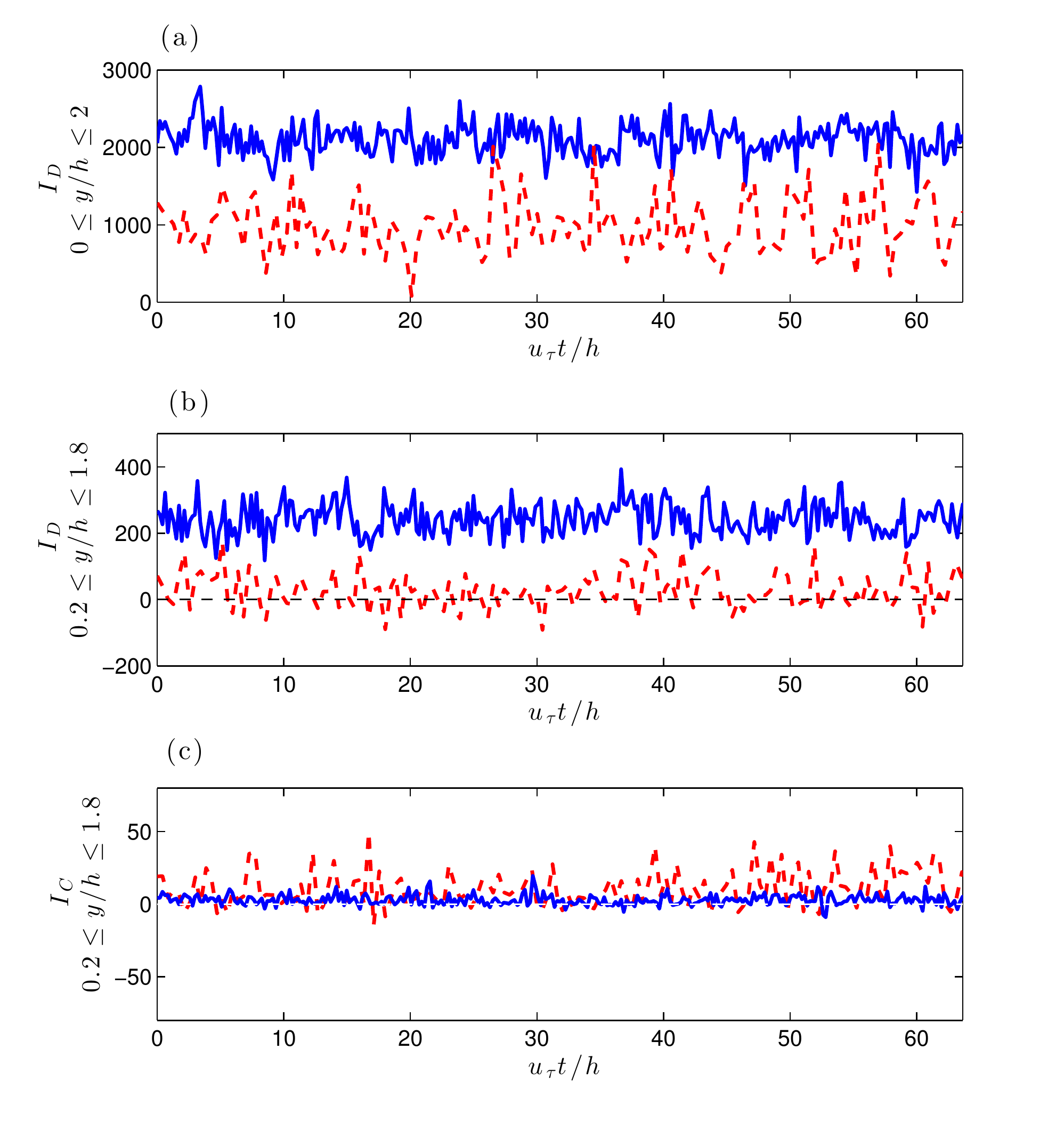}\hspace{.02\textwidth}%
\begin{minipage}{.42\textwidth}\vspace{-9.9cm}\caption{\label{fig:OMx^2}Time series of   the contribution to the time rate of change of $\int\df y\[\Omega_x^2/2\]$ by perturbation torques, $I_D(t)$, and from advection of streamwise mean vorticity by the mean flow, $I_C(t)$, for NS950 (solid) and RNL950 (dashed). (a) $I_D$ for the whole channel, $0\le y/h\le 2$ ($I_C=0$ in this case). The time mean $\widehat I_D$  is $2103.6$ for NS950 and $982.8$ for RNL950. (b) $I_D$ over the outer layer, $0.2\le y/h\le 1.8$. The  time mean $\widehat I_D$ for this region is $242.5$ for NS950 and only $28.7$ for RNL950. (c) $I_C$ for the outer layer $0.2\le y/h\le 1.8$. The time mean $\widehat I_C$ is $2.9$ for NS950 and $11.2$ for RNL950. These figures show that in NS the roll is maintained locally by the perturbation Reynolds stresses and that in RNL the major contribution to the roll maintenance is contributed  locally from perturbation induced torques.}
\end{minipage}
\end{figure}

From Eq.~\eqref{eq:NSm} we obtain  that $\Omega_x$ satisfies the equation:
\begin{align}
\partial_t \Omega_x &=\underbrace{ -V\,\partial_y \Omega_x-W\,\partial_z \Omega_x  \vphantom{\(\overline{v'^2}\)} }_{\textrm{C}} +   \underbrace{(\partial_{zz}-\partial_{yy}) \( \overline{v'w'}\) -\partial_{yz} \(\overline{w'^2}-\overline{v'^2}\)}_{\textrm{D}}+ \,\nu\, \Omega_x\,\Del \Omega_x~.\label{eq:OMx}
\end{align}
Term~C expresses the streamwise vorticity tendency due to advection of $\Omega_x$ by  the streamwise mean flow $(V,W)$. Because there is no vortex stretching contribution to $\Omega_x$ from the  $(V,W)$ velocity field, this term only advects the $\Omega_x$ field and can not sustain it against dissipation.
However, this term may be responsible for systematic advection of $\Omega_x$  from the inner to the outer layer. Term~D is the torque induced by the perturbation field. This is the only term that can maintain $\Omega_x$.   The overall budget for square streamwise vorticity in the region $R$, $y_1 \le y \le y_2$, $0\le z \le L_z$,  is given by: 
\begin{align}
\partial_t \, \int_{y_1}^{y_2} \df y \[ \frac1{2}\Omega_x^2  \] 
&=  \underbrace{ - \[ \frac1{2}\Omega_x^2\,V\]\left.\vphantom{\frac1{2}}\right|^{y=y_2}_{y=y_1} }_{h^{-2} \ut^3\,I_{C}} + \underbrace{\int_{y_1}^{y_2} \df y\,\[\bit\Omega_x\times\textrm{Term D}\]}_{h^{-2} \ut^3\,I_{D}}+  { \,\nu\int_{y_1}^{y_2} \df y\,\[\vphantom{\dispdot{\Omega}_x^{(\textrm{A})}}\bit\Omega_x\,\Del \Omega_x\]}~.
\label{eq:OMx^2}
\end{align}
where we have defined $I_C (t) = h^{-1}\int_R\df y\;\mathcal{I}_{C}(y,t)$ and $I_D (t) = h^{-1}\int_R\df y\;\mathcal{I}_{D}(y,t)$,  with $\;\mathcal{I}_{C}(y,t) =h^3\ut^{-3}\,L_z^{-1}\int \df z\;\Omega_x \times \textrm{(Term~C)}$ and $\;\mathcal{I}_{D}(y,t) =h^3\ut^{-3}\, L_z^{-1}\int \df z\;\Omega_x \times \textrm{(Term~D)}$. Term $I_C$ represents the flux of vorticity into the region  by the  streamwise averaged wall-normal velocity, $V$.

Time series of the contributions from $I_C(t)$ and $I_D(t)$ to the $\Omega_x$  production for  NS950 and RNL950,  shown in Fig.~\ref{fig:OMx^2}, demonstrate that $\Omega_x$ is primarily generated in situ by Reynolds stress torques.
The corresponding  wall-normal structure of the time mean   $\widehat{\mathcal{I}_D}$, representing the local contribution
to streamwise mean vorticity generation from perturbation Reynolds stress induced torques is shown in  
Fig.~\ref{fig:UsdotyOMxdoty}\hyperref[fig:UsdotyOMxdoty]{c}.  Note that for NS in the outer layer the streamwise mean vorticity generation by the Reynolds stress is strongly positive 
at each instant. This is so despite the fact that the r.m.s. $V$ velocity is smaller in the NS than in the RNL, 
as seen in Fig.~\ref{fig:UyVrmsUsrms}, implying greater streamwise vorticity dissipation in the NS.
It could be argued that the positive  value of the generation $I_D$ found in NS and RNL is  a consequence of the 
finite streamwise extent of the channel and as the channel length increases, assuming that there 
is no systematic correlation between streamwise average torques and streak structure,  $I_D$ 
should decrease as $1/\sqrt{L_x}$  and in the limit of  an infinite channel  the NS and RNL 
should sustain no $\Omega_x$. However, S3T theory shows that 
there is a systematic correlation between streaks and roll generation 
by perturbation torque resulting from the deformation of the turbulence by the streak and as 
a result it predicts that in the limit $L_x \rightarrow \infty$, $I_D$ should asymptote to a finite non-zero value at least in RNL.

Having established that  the streamwise vorticity in the outer layer is generated in situ from local Reynolds stress divergences we conclude that the SSP cycle is operational in the outer layer just as in the inner layer.

\section{Discussion and Conclusions}

We have established that RNL self-sustains turbulence at  moderate Reynolds numbers in pressure driven channel flow, despite its greatly simplified dynamics when compared to NS.
Remarkably, in the RNL system, the turbulent state is maintained by a small set of structures with  low  streamwise  wavenumber Fourier components (at $\Ret=950$ with the chosen channel the SSP involves only the $k_x=0$ streamwise mean and  the next  $6$ streamwise Fourier components). Not only that, but this minimal turbulent dynamics arises spontaneously when the RNL system is initialized by NS turbulence at the same Reynolds number. In this way RNL spontaneously produces a turbulent state of reduced complexity. RNL identifies an exquisitely contrived SSP cycle which  has been previously identified to
 comprise  the generation of the streak  through lift-up by the rolls, the maintenance of the rolls by torques induced by the perturbations which themselves are maintained by an essentially time-dependent parametric non-normal interaction with  the streak (rather than e.g.~inflectional instability of the streak structure)~\citep{Farrell-Ioannou-2012}. The vanishing of the Lyapunov exponent associated with the SSP is indicative of a feedback control process acting between the streaks and the perturbations by which the parametric instability that sustains the perturbations on the time dependent streak is reduced to zero Lyapunov exponent, so that the turbulence neither diverges nor decays.

We have established that both NS and RNL produce a roll/streak structure in the outer layer and that an SSP is operating there despite the low  shear in this region. 
It has been  shown  elsewhere that turbulence self-sustains in the log layer in the absence of boundaries~\cite{Mizuno-Jimenez-2013}. This is consistent  with our finding that an SSP cycle exists in both the inner-layer and outer-layer.

The turbulence maintained in RNL is closely related to its associated NS turbulence  and both exhibit a log layer, although with substantially different von K\'arm\'an constants.
Existence of a log layer is a fundamental requirement of asymptotic matching between regions with different spatial scaling, as was noted by Millikan~\citep{Millikan-1938}.
However, the exact value of the von K\'arm\'an constant does not have a similar fundamental basis in analysis and  RNL turbulence, which is closely related to NS turbulence but more efficient in producing Reynolds stress, maintains  as a consequence a smaller shear and therefore greater von K\'arm\'an constant. Specifically, we have determined that the SSP cycle in RNL is characterized by a more energetic and larger scale   perturbation structure, despite having a lower amplitude streak and mean shear.

Formation of roll/streak structures in the log layer is consistent with the universal mechanism by which turbulence is modified by the presence of a streak in such way as to induce growth of a roll structure configured to lead to continued growth of the original  streak. This growth process underlies the non-normal parametric mechanism of maintaining the perturbation variance in the SSP that maintains turbulence if the Reynolds number is large enough~\citep{Farrell-Ioannou-2012}.  
This universal mechanism  does not predict nor require that the roll/streak structures be of   finite  streamwise extent 
and in its simplest form it has been demonstrated that it supports roll/streak structures with zero streamwise wavenumber. From this point of view the observed length of roll/streak structures is not  a consequence of the primary mechanism of the SSP  supporting them but rather a secondary effect of  disruption by the  turbulence.
In this work we have provided evidence  that NS turbulence is persuasively related in its dynamics  to RNL turbulence. 
Moreover, given that the dynamics of RNL turbulence can be understood fundamentally from its direct relation with S3T turbulence we conclude  that the mechanism of turbulence in wall bounded shear flow is the roll/streak/perturbation SSP that was previously identified to maintain S3T turbulence.

\ack
This work was funded in part by the Multiflow program of the European Research Council. Navid Constantinou acknowledges the support of  the Alexander S. Onassis Public Benefit Foundation. Brian Farrell was supported by  NSF AGS-1246929.
We thank Dennice Gayme for helpful reviewing comments.

%
%

\providecommand{\newblock}{}

\end{document}